\documentclass[acmtog, nonacm]{acmart}

\usepackage{wrapfig}
\usepackage{stfloats}

\usepackage{booktabs}
\usepackage{units}

\usepackage{overpic}

\newcommand{\secref}[1]{Sec.~\ref{#1}}

\newcommand{\figref}[1]{Fig.~\ref{#1}}

\usepackage{amsfonts}
\usepackage{amsmath}
\usepackage{caption}

\usepackage{dsfont}
\newcommand{\R}{\mathds{R}}

\newcommand{\replace}[2]{#2}
\newcommand{\change}[1]{#1}
\newcommand{\rev}[1]{#1}

\usepackage[ruled]{algorithm2e} 

\SetAlFnt{\small}
\SetAlCapFnt{\small}
\SetAlCapNameFnt{\small}
\SetAlCapHSkip{0pt}

\usepackage{enumitem}

\acmBooktitle{}

\begin{document}


\title{Designing Personalized Garments with Body Movement}

\author{Katja Wolff}
\email{katja.wolff@inf.ethz.ch}
\author{Philipp Herholz}
\email{ph.herholz@gmail.com}
\affiliation{%
  \institution{ETH Zurich}
  \country{Switzerland}
}
\author{Verena Ziegler}
\email{ziegler@opendress.com}
\author{Frauke Link}
\email{link@opendress.com}
\author{Nico Brügel}
\email{bruegel@opendress.com}
\affiliation{%
  \institution{OpenDress GmbH}
  \country{Germany}
}
\author{Olga Sorkine-Hornung}
\email{sorkine@inf.ethz.ch}
\affiliation{%
  \institution{ETH Zurich}
  \country{Switzerland}
}

\begin{abstract}
	The standardized sizes used in the garment industry do not cover the range of individual differences in body shape for most people, leading to ill-fitting clothes, high return rates and overproduction. Recent research efforts in both industry and academia therefore focus on virtual try-on and on-demand fabrication of individually fitting garments. We propose an interactive design tool for creating custom-fit garments based on 3D body scans of the intended wearer. Our method explicitly incorporates transitions between various body poses to ensure a better fit and  freedom of movement. The core of our method focuses on tools to create a 3D garment shape directly on an avatar without an underlying sewing pattern, and on the adjustment of that garment's rest shape while interpolating and moving through the different input poses. We alternate between cloth simulation   and rest shape adjustment  based on stretch to achieve the final shape of the garment. At any step in the real-time process, we allow for interactive changes to the garment. Once the garment shape is finalized for production, established techniques can be used to parameterize it into a 2D sewing pattern or transform it into a knitting pattern.
\end{abstract}

%
%

\ccsdesc[500]{Computing methodologies~Computer graphics}
\ccsdesc[500]{Computing methodologies~Shape modeling}
\ccsdesc[500]{Computing methodologies~Mesh geometry models}
\ccsdesc[300]{Computing methodologies~Digital Garments}

%
%

\keywords{computational fabrication, garment modeling, cloth simulation}

\begin{teaserfigure}
	\centering
	\setlength{\tabcolsep}{2pt}
	\begin{overpic}[grid=false,width=\linewidth]{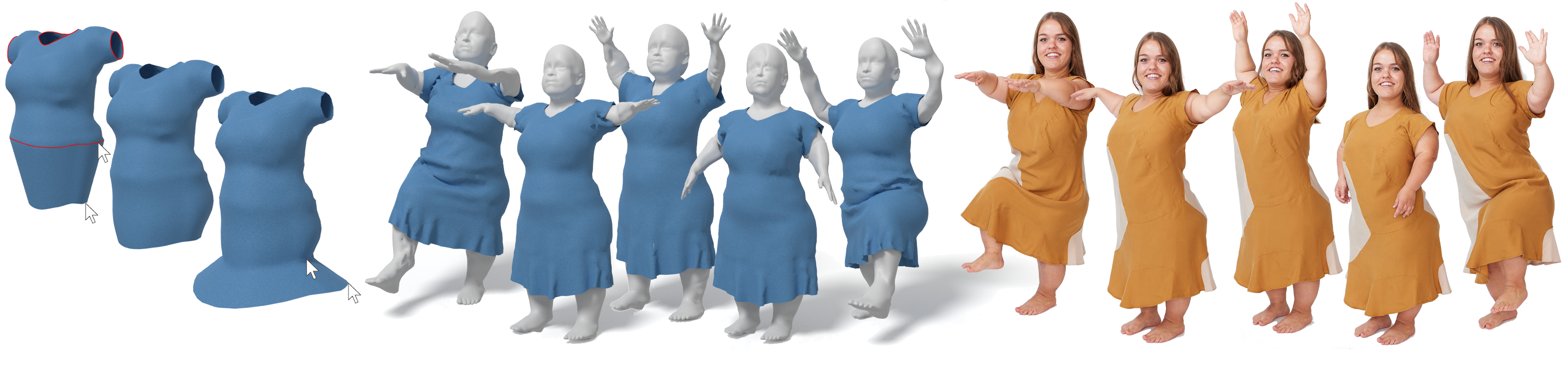}
	    \put(5,0){(a) garment design}
	    \put(29,0){(b) garment simulation \& adaption}
	    \put(75, 0){(c) physical garment}
    \end{overpic}
    	\caption{
	Our method allows casual users to interactively design custom garments.
    We provide six intuitive tools to design and adjust the initial garment shape on a 3D avatar generated by scanning the user in different poses.
	We then simulate the garment interacting with the animated avatar. Throughout the process, we adjust the garment model to achieve the final garment shape, such that it comfortably fits in all poses.
	We verify our method by physically fitting sewn garments on the user. Our method is able to create garments for people who fall far outside the range of standard sizes, like this adult, whose height is \unit[125]{cm}. Please refer to the accompanying video for a more detailed view of the dress.}
	\label{fig:results_sofia}
\end{teaserfigure}

\maketitle

\begin{figure}[t]
	\centering
	\includegraphics[width=\linewidth]{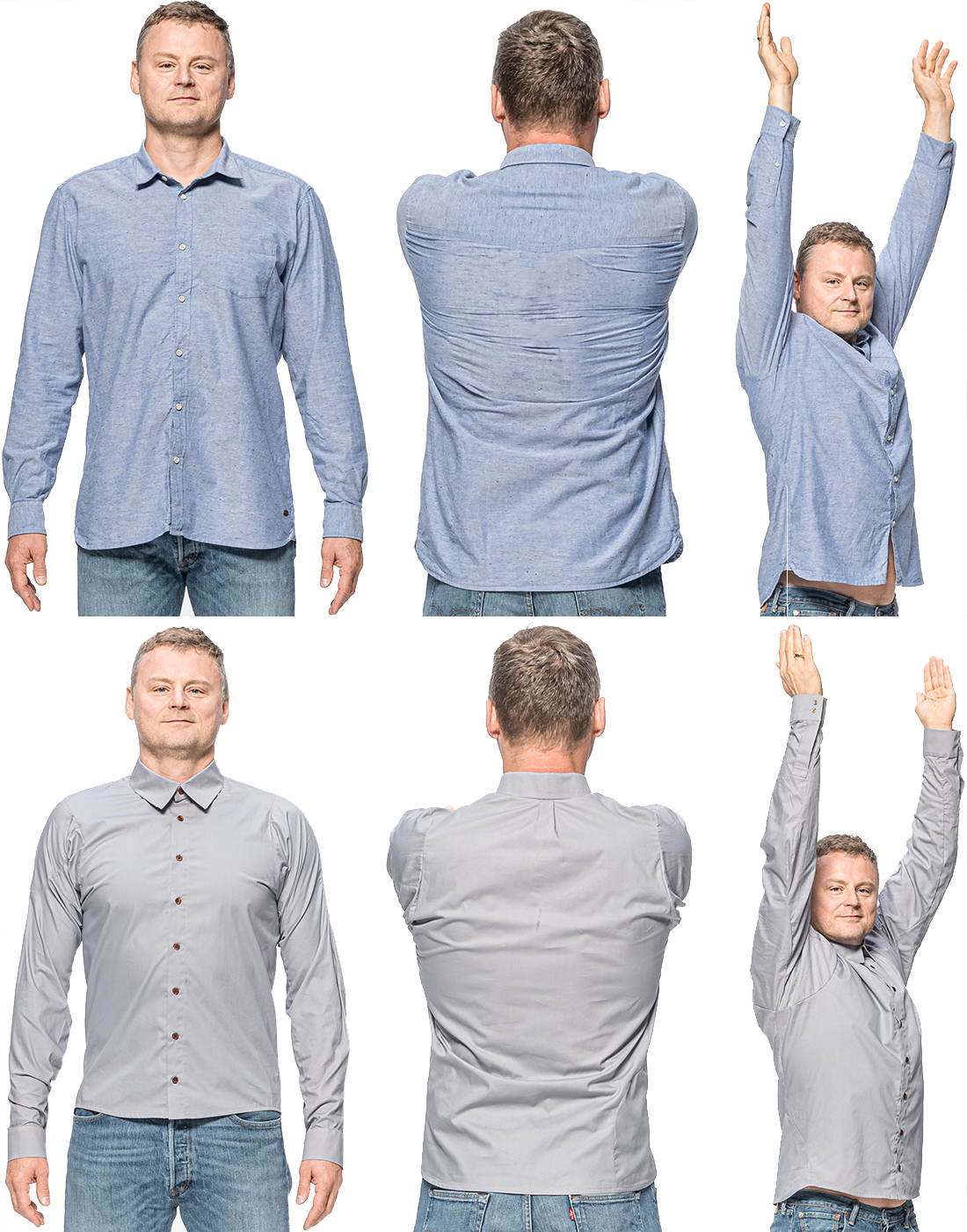}
	\caption{We compare a standard shirt of size 43, modern fit (top row) with our custom-fit shirt (bottom row). Our shirt fits more tightly to the body (left), while allowing more comfort in a range of motions (middle and right). The standard shirt stretches and tightens uncomfortably when the arms are outstretched to the front (middle). Lifting the arms leads to pulled down sleeves and uncomfortable stretch along the arms of the standard shirt  (right). See \figref{fig:results_shirt} for more details of our custom-fit shirt.}
	\label{fig:compare_shirts}
\end{figure}

\section{Introduction}

The garment industry is a trillion-dollar, global industry that uses a large amount of natural and human resources \cite{bick2018global}. According to recent research, of all annually manufactured garments, 25\% are never sold, and another 25\% are sold but almost never worn \cite{morlet2017new}, meaning that nearly half of the produced items are imminently destined for landfill or incineration. One reason is that for many people it can be challenging to find fitting clothes, as standardized sizes often cannot account for individual differences, such as longer or shorter arms, asymmetries, body dimensions that fall outside the commercially available range or even missing extremities. Standard sizes available in stores vary over the globe; in Germany, 70\% of women do not fit the commonly available standard sizes \cite{SizeGermany}. Moreover, people who share the same standard measurements  might still vary drastically in their body shape. All these issues lead to high return rates of purchased garments. In recent years, on-demand, individualized fashion has become a focus in research and industry as a response to these problems. Still, the garment industry so far remains largely unchanged and depends on strenuous manual labor and mass-produced ready-to-wear garments. Fitting garments individually is currently a manual and expensive process done by professional tailors, who adjust existing sewing patterns to the person's size and features while leaving the overall pattern design unchanged. This process did not substantially change for centuries. Even modern computer based garment design systems, like CLO 3D \cite{CLO3D}, still rely on traditional 2D sewing patterns as their central design space.

We take a \replace{radically}{} different design approach, which is liberated from 2D sewing patterns and focuses on maximum fit and comfort under a range of individual motions for any kind of body shape. Our method creates an optimized shape of the garment, and only in a second step prepares it for production using existing techniques. We create \replace{completely}{} new sewing pattern designs via 2D parametrization. Alternatively seamless knitting patterns can be generated based on our designs, sidestepping the use of 2D sewing patterns completely. We do not use traditional sewing pattern designs and symmetry \rev{to accommodate all body shapes. Though many traditional garments are symmetric, modern fashion designers (like Shingo Sato) or trends like color blocking for dresses creatively break free from these limitations as well. Once designed, irregular patterns are not difficult to cut or sew, as confirmed by the professional tailors we work with. Nevertheless we }still allow the traditional positioning of seams if desired.

We work with a variety of poses for each person: we scan each pose using a commercial 3D scanner and use existing methods to register a template body mesh to each scanned pose, such that we obtain pose meshes of matching connectivity. This provides us with higher fidelity to the true shape of the user's body than simply using a few standard measurements and a pre-set, symmetric avatar, and also helps eliminating measurement errors by the user. For each garment we use a select number of poses that this garment needs to accommodate. We provide a set of intuitive tools to design the garment shape directly in 3D, sidestepping the necessity for creating an underlying sewing pattern. We enable the drawing of garment boundaries directly on the avatar to create skintight clothing, but also allow for the addition of loose parts. An optional paintbrush tool can be used to add cloth in specified regions, and by defining a minimum distance of the garment to the body, we can explicitly control comfort. The garment is then simulated using existing cloth simulation techniques. We maintain a \emph{rest shape}, which is the garment without any forces applied and a \emph{simulation shape} which represents the current simulation mesh undergoing stretching, bending and shearing as it is deformed by the dynamic body mesh. While smoothly transitioning between different poses, we compute a stretch metric of the garment and adjust the garment rest shape whenever the stretch exceeds a threshold. This process is fast and allows real-time interaction to adjust the garment at any point in the process. The rest shape can then be used for production, e.g., by applying existing methods to compute a distortion-minimizing sewing pattern. The resulting garments fit more tightly, while allowing for a wider range of motions compared to garments of standard sizes (see \figref{fig:compare_shirts}).

\subsection{Contributions}

\replace{}{The central contribution of our paper is the formulation of a novel and easy iterative garment design algorithm that uses stretch optimization to take varying body poses and body movements into account.}

Using our software, we create a number of garments for people of widely varying stature and body type and demonstrate professionally manufactured garments based on these designs in Figures \ref{fig:results_sofia}, \ref{fig:compare_shirts}, \ref{fig:results_extraguy}-\ref{fig:results_jumpsuit}, and in the accompanying video. Our source code is available at \url{https://github.com/katjawolff/custom_fit_garments} to foster further research.

\section{Related work}
\label{sec:rel}

Computational garment design has become a highly active research field in different scientific areas over the past years. We concentrate on the most relevant works  in relation to our contribution and group them according to their focus on garment design, fit and simulation. 

\paragraph*{Garment design.}
Works on garment design focus mainly on providing tools for automatically creating and manipulating the garments and their underlying sewing patterns, such that designers can easily and quickly explore design choices. 
Nayak and Padhye~\shortcite{nayak:2017:automation} broadly survey the use of automation in garment manufacturing, including computer aided design. Early works focus on interactive design and modification pipelines, providing visual real time feedback \cite{Keckeisen:TailorTools:2004, Volino:EarlyGarmentDesign:2005}. 
Berthouzoz et al.~\ \shortcite{Berthouzoz:ParsingPatterns:2013} scan and parse existing, traditionally published patterns and convert them into 3D garment models. Umetani et al.\ \shortcite{Umetani:SensitiveCouture:2011} introduce a system for bidirectional interactive garment design that allows to edit both the 2D pattern and the 3D garment shape, while keeping the correspondence between the two. To facilitate fabrication of computed patterns, Igarashi et al.\ \shortcite{Igarashi:SeamAllowance:2008} automatically generate necessary seam allowance for sewing. 
To create seamless print designs, Lu et al.~\shortcite{lu2017new} paint directly on a garment in 3D and transfer the print to the sewing pattern, and Wolff et al.~\shortcite{wolff2019wallpaper,wolff2019reflection} adjust the positioning of pattern pieces on a given textured fabric, while also slightly adjusting the sewing pattern shape.
Several commercial CAD fashion design softwares are available nowadays, including CLO 3D \shortcite{CLO3D} and Optitex \shortcite{optitex}, which enable the digital design of sewing patterns and their draping, greatly accelerating the iterative design process. However, they still follow the traditional design workflow, which requires a professional garment designer with experience in modifying 2D sewing patterns.

A multitude of sketch based methods \cite{Decaudin:VirtualGarments:2006, Turquin:SketchInterface:2007, Rose::DevelopableSurfaces:2007, Robson:ContextAwareGarments:2011} create 3D garment shapes from contours, boundary lines and seam lines that are drawn on a digital model. Since garments are sewn from flat sheets of cloth, developable patches are automatically computed for the sewing pattern. Incorporating advancements in machine learning, Wang et al.\ \shortcite{Wang:GarmentShapeSpace:2018} use a data-driven approach to estimate garment shapes from a sketch of a desired fold pattern. Instead of creating a new garment, the approach by Li et al.\ \shortcite{Li:2018:FoldSketch} enriches a garment with folds and pleats guided by sketches. In order to incorporate the traditional workflow of pattern design while keeping the advantages of working digitally, Wibowo et al.\ \shortcite{Wibowo:Dressup:2012} use a physical real-world mannequin as a guide for drawing with a specialized tool in 3D around it.

Garment design for fabrication or use on digital avatars is commonly based on 2d sewing patterns created by a skilled professional who ensures fit and style. We take a different design approach only focusing on the 3D shape and fit of a garment. This gives us more degrees of freedom at design time.
The resulting shape can then be directly knit \cite{narayanan2019visual} or flattened with minimal distortion by existing parametrization methods, such as \cite{sharp2018variational} \change{or concurrent work by Pietroni et al.\shortcite{pietroni2022computational}}. 

The resulting garments show a novel and intriguing style and differ significantly from traditional sewing patterns. Similarly, Kwok et al.\ \shortcite{kwok:2016:styling} create styling curves directly on the avatar to create novel sewing pattern designs for sports garments, but in their work, the fit is simply assumed to exist and is not optimized. 

\paragraph*{Garment fitting.}

A number of works address custom-fitting garments for different body sizes and shapes.
Early research focuses on fitting the 3D shape of a pre-designed template garment to an arbitrarily sized avatar by creating feature correspondences between the avatar and the garment \change{ \cite{cordier2003made,wang2005design,meng2012flexible,meng2012computer}}. The 2D sewing patterns are \change{either }only created afterwards through different parametrization methods \cite{wang2005design,meng2012flexible,Decaudin:VirtualGarments:2006}\change{, or adjusted through a mapping to selected style lines in 3D \cite{meng2012computer}}. The seams that define these pieces are either sketched \cite{wang2005design, Decaudin:VirtualGarments:2006} or transferred from the initial input garment \change{\cite{meng2012flexible, meng2012computer}}. 
Introducing style and fit criteria, like proportion, scale, shape and fit, allows Brouet et al.\ \shortcite{Brouet:DesignPresGarTrans:2012} to grade existing sewing patterns for largely different body sizes. All these methods suffer from the so called draping effect: Since the 2D sewing patterns are directly parameterized pieces of the 3D garment, the resulting sewn garment deforms further when draped. \replace{}{In contrast, we directly modify the rest shape of the garment and allow for an optimized placement of seams in a post process}.
More recent methods adjust for the draping effect. Bartle et al.\ \ \shortcite{Bartle:PhysicsPatternAdj:2016} allow users to mix existing garment designs and calculate the sewing pattern inversely from the garment. Wang\ \shortcite{wang2018rule} solves the garment shape and sewing pattern design as a single nonlinear optimization problem. This method only allows for small changes in avatar size, which are created by applying measurements to deform a base mesh; the method is demonstrated by sewing standard sized garments. In contrast, we incorporate 3D scans of extremely different body shapes to create our garments. Unlike both above methods, our work does not rely on pre-designed sewing patterns and incorporates movement of the body. By working with a 3D rest shape of the garment and optimizing the placement of seams, we also limit the draping effect, allowing us to to incorporate large deformations and asymmetries for custom-made apparel.
The work by Montes et al.~\shortcite{montes2020computational} explores the automatic generation of new sewing patterns. They embed the cloth as a two-dimensional elastic membrane in the surface of an elastic body mesh and adjust an initial 2D pattern to create skintight clothing. Their method allows to optimize the layout for multiple poses simultaneously. \replace{
We also create novel sewing patterns for multiple poses, but in contrast, we do not rely on an initial 2D pattern, and our garment does not need to be skintight and allows wrinkles.}{However, important classes of garments like skirts or loose fitting garments as well as wrinkles cannot be handled by this method.}

\change{The work by Liu et al.~\shortcite{liu2021knitting} provides the tools to create knitwear garments and allows for different motions by adjusting the local knit properties instead of the shape. In contrast, our approach is not limited to knit, but works on woven fabrics as well and adjusts the garment shape instead.}
Notably few previous works demonstrate fit with actual sewn garments, and when they do, it is mostly on \change{mannequins\cite{Decaudin:VirtualGarments:2006, Brouet:DesignPresGarTrans:2012, Bartle:PhysicsPatternAdj:2016,liu2021knitting}}, with the exception of Wang \shortcite{wang2018rule}, who demonstrates the fit of the garments on real humans, as we do.

An emerging body of works employ learning based approaches to automatically dress different body shapes, e.g.  \cite{guan2012drape,pons2017clothcap}. Such methods are mostly targeted at purely virtual try-on and visual appeal, where the resulting garments are not guaranteed to be physically plausible. Unlike our work, these works do not yet address physical fit and fabrication.

\paragraph*{Cloth simulation.}
The simulation of cloth has been of interest for decades with different foci, approaches and models, depending on whether an application requires speed or physical accuracy and the type of cloth (knit or woven). Cloth elasticity can be simulated based on fast and simple spring models \cite{choi2005stable, liu2013fast}, on continuum models \cite{baraff1998large, narain2012adaptive}, which can work well for woven cloth, or the individual yarns \cite{cirio2014yarn, kaldor2010efficient}, which is very useful in modeling intricate knit patterns or a snag of a single thread. Research has also covered aspects such as collision handling \cite{bridson2005simulation, tang2018cloth}, measurement of cloth elasticity \cite{wang2011data, miguel2012data}, inextensibility \cite{goldenthal2007efficient} or plastic deformations \cite{jung2016modeling}. Though our framework is relatively independent of the particular choice of the simulation method, we do substantially rely on cloth simulation and base our method on the model by Baraff and Witkin\ \shortcite{baraff1998large}. The different components of the energy, such as stretch, bend or shear, can be individually adjusted \cite{bergou2006quadratic, tamstorf2013discrete}.

\paragraph*{Human body modeling.}

Our method is targeted at creating garments for real people. To capture details of individual body shapes, we therefore rely on the large body of research on object scanning and body shape modeling, as well as commercially available 3D scanners. Several methods for fitting a parameterized, articulated  human body model to 3D scans exist, such as the multi-person linear model SMPL, based on skinning and blend shapes and learned from thousands of 3D body scans \cite{SMPL:2015}, the more recent STAR model \cite{osman2020star}, as well as dynamic models like Dyna \cite{Dyna:SIGGRAPH:2015}. The FAUST dataset \cite{bogo2014faust} allows to evaluate and compare body models, and we use several avatars of this dataset to demonstrate our method. Our framework is independent of the chosen scanning and registration technique, which can be replaced once even more accurate future methods become available. 


\section{Method}
\subsection{Overview}
Mastering garment design and manufacturing 
requires years of education and experience. Central challenges are the creation or variation of 2D cut patterns and optimizing fit in different poses.
Our goal is to design a system that assists casual users that do not have the required expertise in creating personalized garments. To this end, we wish to enable users to  express their artistic intent without the need to handle technicalities like cut patterns or reasoning about fit in different poses. Our system is built around two key ideas that allow us to reach this goal.

Our method is conceptually simple and we rely on established geometry processing algorithms to implement it.

\paragraph*{3D garment design.}
Instead of working with 2D cut patterns, the user is creating the garment in 3D using an intuitive set of tools. The resulting 3D garment can then either be produced using knitting or traditional sewing. For the latter technique we automatically generate cut patterns from 3D garment geometry using an advanced parameterization technique \change{\cite{sharp2018variational}} (\secref{sec:fabrication}). 
Our user interface is centered on a scanned 3D avatar of the intended wearer in different poses. This scan has to be created only once and can be reused to design different garments. In \secref{sec:poses} we present details about how to create the personal avatar.

\paragraph*{Optimizing fit.}
Since the term ``fit'' can be quite subjective, we first quantify what our system considers to be a good fit. A central aspect is how well the garment is adapting to the body. A user might want a garment that sits tight on the body in some areas while being  looser in others. We provide a simple 3D painting interface that allows the user to indicate where more fabric should be added, resulting in a looser fit. 
Beyond this simple definition of fit, we take \emph{dynamic} fit into account. Garments are often designed for comfortable wear in a neutral pose and may restrict the range of motion. A shirt, for example, commonly restricts the movement of the arms, limiting the space that can be reached in order to achieve a tighter fit. Sportswear should be optimized to support sport-specific  movements without tearing the garment.
\change{Therefore, for each type of garment we choose a set of poses that constitutes the maximum movement the garment needs to accommodate.}
We measure dynamic fit as the maximum stretch a garment has to endure while moving through the predefined set of body motions. We  define how we measure stretch in \secref{sec:rest_shape2}. 
In order to optimize dynamic fit we could let the user introduce extra cloth manually and loosen the fit at places where  issues might occur. However, identifying the locations where dynamic stretch occurs is non-trivial. To keep our system simple and accessible we opt for an automatic strategy based on cloth simulation together with a custom adaption strategy that reduces stretch by modifying the garment's rest shape (\secref{sec:rest_shape2}). 

\subsection{Body pose acquisition and interpolation}
\label{sec:poses}

\begin{figure}[t]
	\centering
	\begin{overpic}[width=\linewidth]{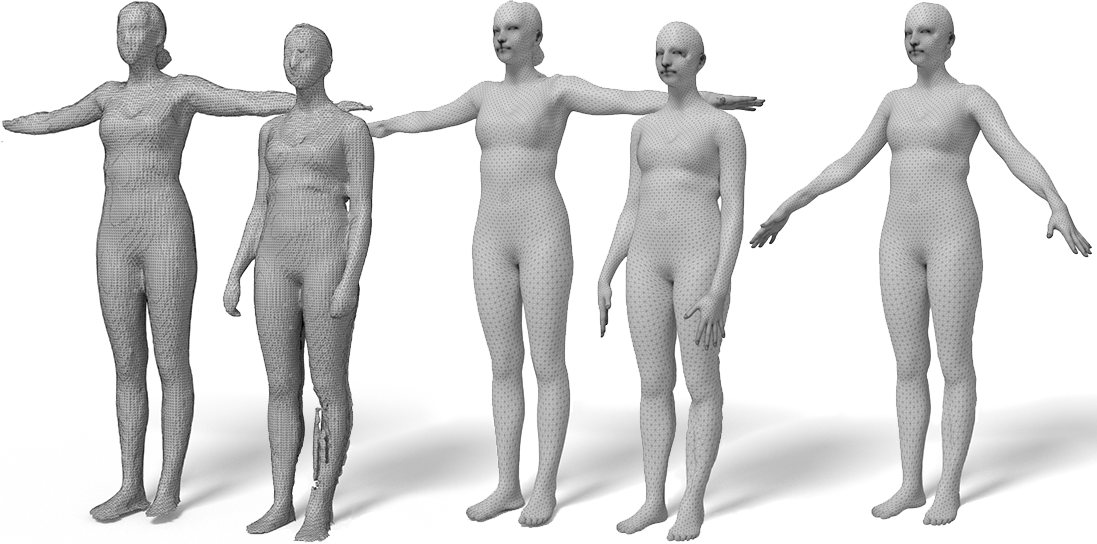}
	\put(32,0){\small (a)}
	\put(66,0){\small (b)}
	\put(90,0){\small (c)}
	\end{overpic}
	\caption{Using a 3D scanner, we obtain a mesh for each pose (a) and we register all the meshes based on the SMPL model \cite{SMPL:2015} (b). To create smooth motions between poses, we compute interpolated in-between meshes (c).}
	\label{fig:body_poses}
\end{figure}

\paragraph*{Acquisition.} In order to incorporate the movement of the human body in the garment optimization, we capture different static poses of the individual. The amount and nature of these poses depends on the desired garment. A comfortably fitting garment designed for sports requires a wider range of poses than a tight fitting cocktail dress. A long-sleeved shirt might need two poses to capture bent and outstretched elbows, as opposed to a short sleeved garment. The specific technique used to acquire the body poses can be chosen freely, as long as all poses are represented as meshes with the same number of vertices and connectivity in full correspondence across the different poses. We use a \emph{Structure Sensor} \cite{occipital} to scan each pose. The resulting mesh can include background objects, which we remove, as well as holes and artifacts. Two cleaned up meshes are shown in  \figref{fig:body_poses} (a). In order to create evenly meshed poses in full correspondence, we use \emph{Meshcapade} \cite{meshcapade} which employs the SMPL model \cite{SMPL:2015} (\figref{fig:body_poses} (b)). These technologies can be easily substituted by future developments.  

\paragraph*{Interpolation.} 
During the later garment simulation, we want to smoothly transition between the captured poses in order to optimize the garment for a range of motions. Given the captured pose meshes with corresponding vertices and faces, we can readily interpolate these poses. Even though Meshcapade provides skeletons with each registered pose, we do not use those for interpolation, for two reasons: (i) We want our algorithm to be independent of the registration technique and work for methods that do not supply a skeleton, and (ii) we wish to be able to incorporate changes in the body shape that are not necessarily captured by skeletal mesh deformations (skinning), such as movement of fat and tissues, or volumetric changes for the same pose (e.g., growing belly in  pregnancy). Therefore, we directly interpolate the pose meshes using a nonlinear morphing technique akin to deformation transfer \cite{sumner2004deformation} and Poisson shape interpolation \cite{xu2006poisson}. Though in theory this approach can create self-intersections, we did not observe any in our examples. This step could be replaced by more advanced approaches in the future based on e.g. muscle simulation.

\begin{figure*}[t]
	\centering
	\begin{overpic}[width=\linewidth]{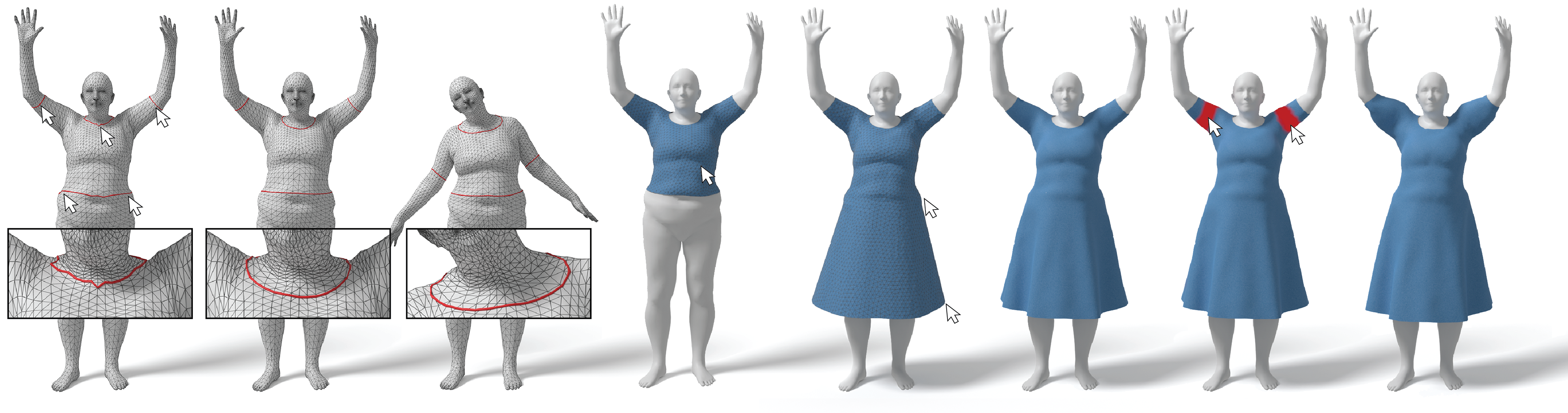}
	\put(5,0){\small (a)}
	\put(17.5,0){\small (b)}
	\put(30,0){\small (c)}
	\put(42,0){\small (d)}
	\put(55,0){\small (e)}
	\put(67,0){\small (f)}
	\put(78.,0){\small (g)}
	\put(90.25,0){\small (h)}	
	\end{overpic}
	\caption{Our toolset allows creating garment boundaries on the avatar mesh by connecting clicked vertices on a shortest path (a). These boundaries are then smoothed (b) and expressed in barycentric coordinates on the triangle mesh, which allows us to transfer them to different poses (c). Clicking an area on the avatar mesh creates a garment mesh enclosed by the created boundaries (d). Garments can be extended with loose-fitting parts at a chosen boundary (e). Even during simulation (f), changes to the garment are possible. We allow painting areas where cloth should be added (g,h). All interactions are marked by a cursor.}
	\label{fig:tools}
\end{figure*}

\begin{figure}[t]
	\centering
	\begin{overpic}[width=\linewidth]{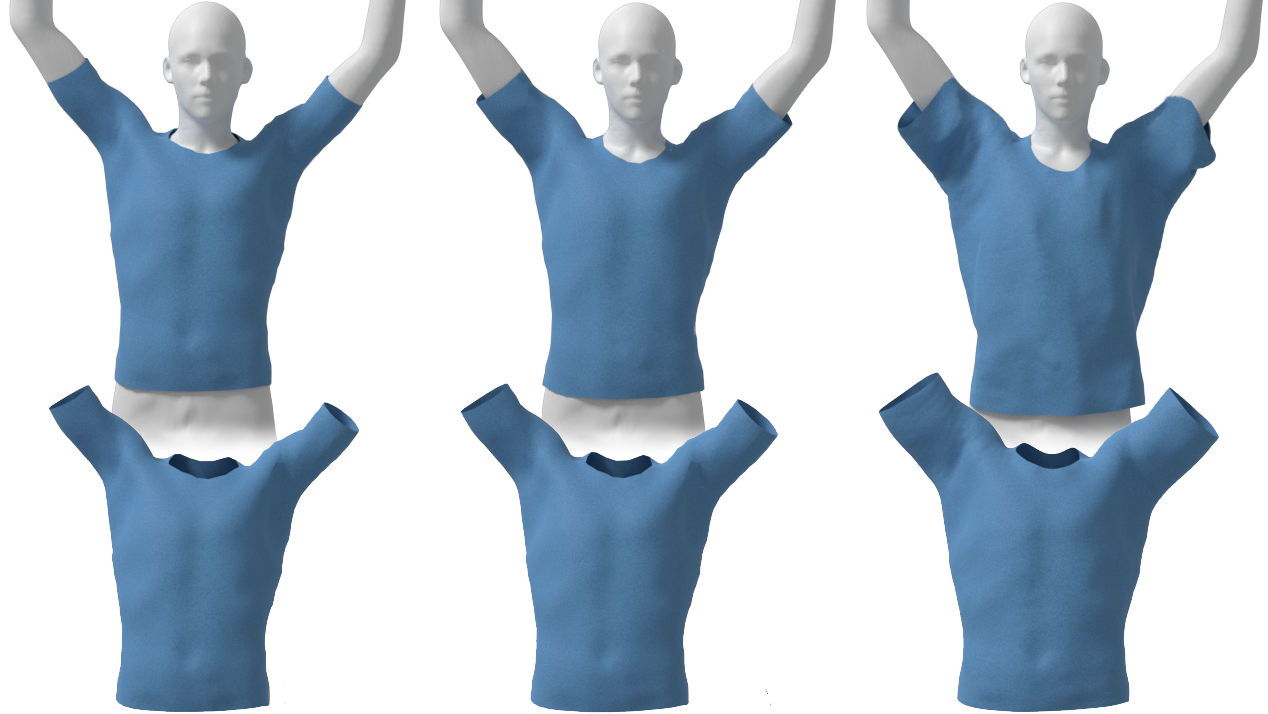}
	\put(22.5,2){\small 0cm}
	\put(56,2){\small 1cm}
	\put(90.5,2){\small 3cm}
	\end{overpic}
	\caption{Different offsets for comfort control. The top row shows the simulated garment shape after no offset or an offset of 1 or 3 cm was applied. The bottom row shows the adjusted rest shape.}
	\label{fig:offset}
\end{figure}

\begin{figure}[t]
	\centering
	\begin{overpic}[width=\linewidth]{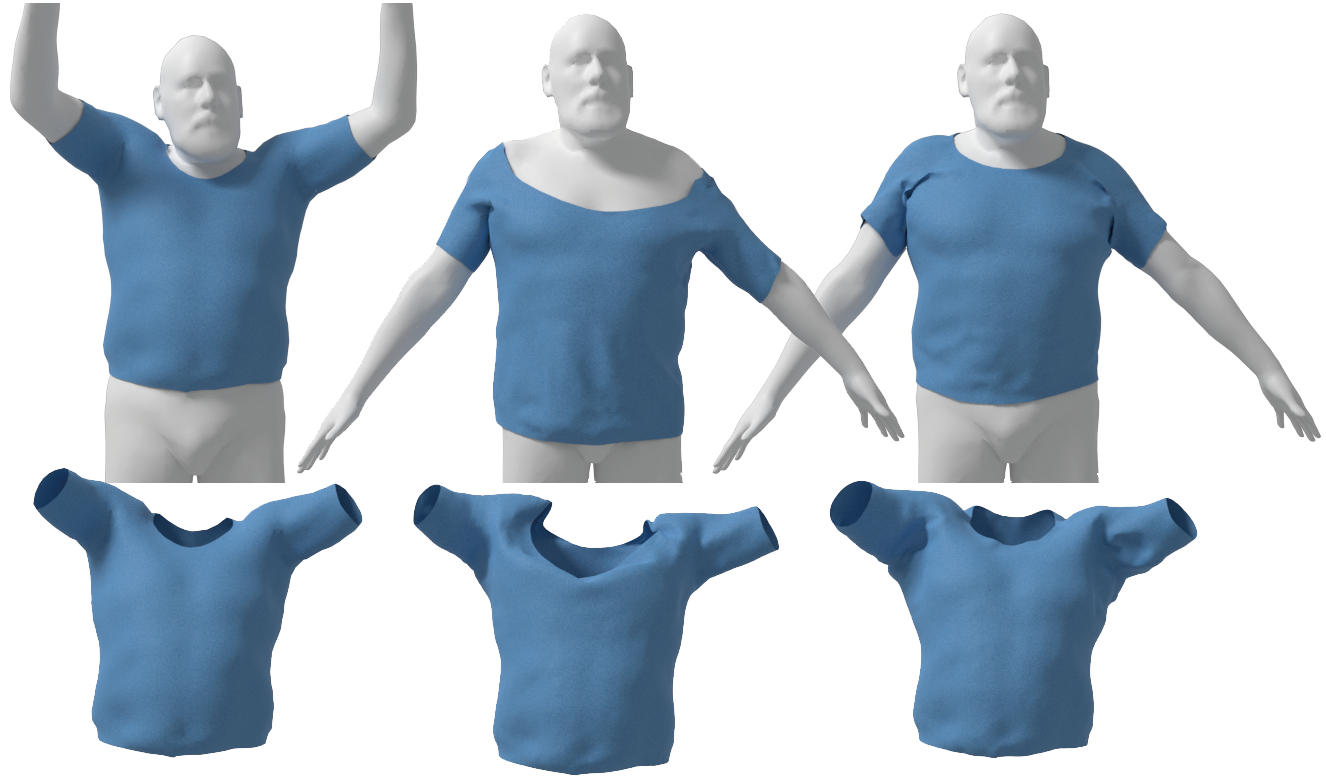}
	\put(23.5,3){\small (a)}
	\put(54,3){\small (b)}
	\put(85,3){\small (c)}
	\end{overpic}
	\caption{When boundaries are not pinned to the avatar while changing poses, the large stretch forces can result in widened hems, such as a wide neckline (b). Pinned boundaries ensure that the garment stays in place and the stretch forces result in enlarged cloth area instead (here in the shoulder regions) to allow movement of the specific body parts (c).}
	\label{fig:fixed_boundaries}
\end{figure}

\subsection{Digital garment creation}
\label{sec:tools}
Starting with any of the captured poses, we use it to design the garment directly in 3D. We propose a set of simple \replace{but powerful }{}tools that allow to explore the complex space of garment designs:

\begin{enumerate}[leftmargin=3\parindent]
    \item[\textbf{(T1)}] \textbf{Boundary tool} to draw garment boundaries on the avatar.
    \item[\textbf{(T2)}] \textbf{Extension tool} for garment parts that do not follow the body shape, like skirts. 
    \item[\textbf{(T3)}] \textbf{Paint tool} to add/remove cloth in specific areas for a looser or tighter fit.
    \item[\textbf{(T4)}] \textbf{Comfort tool} to set a minimum distance to the body.
    \item[\textbf{(T5)}] \textbf{Pinning tool} for fixing garment vertices.
    \item[\textbf{(T6)}] \textbf{Seam tool} to predefine garment seams.
\end{enumerate}
\change{These tools allow the creation of common garments to demonstrate the capabilities of our pipeline, but are not complete and can be extended to allow further designs. Examples of currently unsupported designs are hoods, or non-manifold seams for multiple layers of cloth. Small additions like pockets, cuffs or collars need to be added manually.} 

\paragraph*{Boundary tool.} We allow the user to draw closed loop boundaries directly on the avatar by consecutively clicking points on the mesh, which we connect by shortest edge paths. After the loop is closed, we perform a smoothing operation, which creates a polyline that is defined on the mesh through barycentric coordinates. Since the different poses have a corresponding mesh structure, the defined boundaries are valid for all poses (\figref{fig:tools} (a)-(c)). We can create a garment from these boundaries by specifying a boundary-enclosed region on the avatar. To define the initial garment rest shape we duplicate the enclosed submesh and remesh it to a desired resolution, generating triangles with similar area (\figref{fig:tools} (d)). This tool is used to create a new garment shape, whereas all following tools are used to edit existing garment shapes.

\setlength{\intextsep}{5pt}%
\setlength{\columnsep}{8pt}%
\begin{wrapfigure}{r}{0.25\linewidth}
\centering
\includegraphics[width=\linewidth]{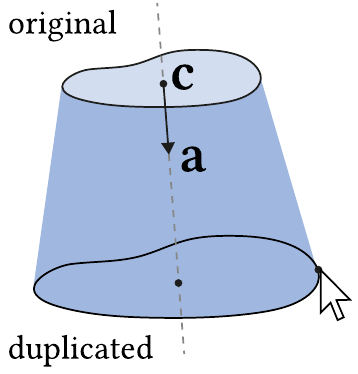}
\end{wrapfigure}
\paragraph*{Extension tool.} In order to allow for the design of dresses, skirts, tops with wide sleeves and similar features that are not skintight, we allow to extend a garment at a chosen boundary. 
An axis is created from this boundary by calculating its center of mass $\mathbf{c}$ and its vector area $\mathbf{a}$. 
The vector area is a vector that is well defined by the vertex positions of the boundary and points in the direction that maximizes the enclosed area when projected onto a plane, and therefore can also be calculated for non-planar boundary loops.
The chosen boundary is duplicated, translated and scaled along this axis, such that it lies on an additionally specified point, which can be defined with a mouse click by the user. The duplicated boundary is then connected to the existing garment mesh. A remeshing operation ensures even meshing. See  \figref{fig:tools} (e).

\paragraph*{Paint tool.} This tool can be used to specify areas that need to be enlarged. The intensity of the color per triangle defines a scaling factor. The garment mesh is then adjusted accordingly by applying the methods described in \secref{sec:rest_shape2}. See \figref{fig:tools} (g)-(h).

\paragraph*{Comfort tool.} We allow the user to set a minimum offset distance between the garment and the body. When the garment is simulated (\secref{sec:simulation}), a simple collision detection with the body pushes the garment vertices away from the body by the set distance and creates a small stretch everywhere. The rest shape adjustment step (\secref{sec:rest_shape2}) then automatically adjusts the garment to counteract this stretch. A small offset is useful to allow for thick textiles or to compensate small errors introduced through the 3D scanning process. Large offsets can also be used as a design choice. See \figref{fig:offset}.

\paragraph*{Pinning tool.} We allow the user to pin selected vertices of the garment mesh to the body during the cloth simulation, specifically whole garment boundaries. The pinning is implemented as an additional constraint (\secref{sec:simulation}). This tool is especially useful when the avatar moves through different poses. As illustrated in \figref{fig:fixed_boundaries}, a garment's neckline might stretch when the avatar lowers its arms. Pinning the neckline prevents this enlargement and keeps the neckline of the rest shape in place. 

\paragraph*{Seam tool.} After the final garment shape is computed by moving through several poses and ranges of motion, we allow to optionally predefine seams by creating boundaries, similar to the \textit{Boundary tool}, but directly on the garment rest shape. This is not necessary, but might be desired to give the garment a certain look, e.g., by defining traditional shoulder seams for men's shirts (\figref{fig:results_shirt_tpose}, \ref{fig:results_shirt}, \ref{fig:results_extraguy}) or by creating seams between regions with different textiles (\figref{fig:results_extragirl}).

\subsubsection{Example modeling session}
\figref{fig:tools} shows a typical modeling session using our set of tools. The user wants to create a dress and starts of   f with a static pose of their choice. First the upper part of the dress is created by defining boundaries on the mesh. To this end, the user selects points on the avatar that can be interpolated by the software to automatically generates smooth boundary curves (b). In case points can not easily be selected, it is possible to change the pose temporarily, the boundary will be transferred to the new pose (c). To generate the initial garment, the user selects the torso and the software will automatically generate cloth inside the region enclosed by the boundaries. Using the extension tool it is easy to extend the shirt into a dress (e). At this point the user decides to run physical garment simulation to inspect how the garment will behave under gravity (f). Since the dress fits very tightly on the arms, the user decides to use some extra cloth to create puffy sleeves. When the user is satisfied with their design, the automatic pose optimization steps generates a garment that fits under movement.


\subsection{Garment simulation \& adaption}
\label{sec:simulation}
To optimize dynamic fit, we want to modify the garment geometry slightly in an attempt to reduce stretch under motion. In this section we will give some details about the simulation method with a focus on how stretch is measured. In the next section we will use these insights to optimize the garment itself.

\subsubsection{Triangle meshes}
Throughout our pipeline we work with triangle meshes. A triangle mesh is given by a set of vertices and triangles. Vertices are represented by points $\mathbf x_i$ in 3D  with $\mathbf x_i \in \mathds R^3$ and $i = 1,\dots, n$.
 Triangles are given by three indices $\mathbf t = \begin{pmatrix} t_0, t_1, t_2\end{pmatrix} \in \left[1,..,n \right]^3$, referencing the vertices that are part of the triangle. If we are interested in the  points of a triangle $\mathbf t$, we use the notation
\begin{align*}
 \mathbf{x}_t = \begin{pmatrix}  \mathbf{x}_{t_0} &   \mathbf{x}_{t_1} &  \mathbf{x}_{t_2} \end{pmatrix} \in \R^{3 \times 3}.
\end{align*}

For the simulation we distinguish between two triangle meshes. The \emph{rest shape mesh} represents the garment that has been modeled by the user. All vertex positions of the rest shape are marked by the \emph{hat} symbol $\hat{\mathbf x}_i$.
When simulating the garment, it gets deformed due to internal forces like stretch, shear, bending and external forces like gravity and collisions with the avatar. The current state of the simulation is called \emph{simulation mesh}. Vertex positions of the simulation mesh are denoted by $\mathbf x_i$. Both meshes share the same triangle set $\mathcal T$.

\subsubsection{Simulation model}
We use the cloth model introduced in \cite{baraff1998large, bridson2005simulation} which has seen wide adoption in research and industry over the years. The physical behaviour of cloth can be described as a material that resists stretch and, to a lesser degree, shearing and bending. How strong cloth resists these forces depends on the specific fabric. 
Like many physically based simulation approaches, the model is based on an energy which sums up per-triangle and per-edge contributions. 

Cloth can have very different material properties in terms of stiffness and bending resistance. By using weighting factors for the stretch, shear and bending part of the energy, we are able to model different material behaviours. However, for all examples we fixed the material parameters to resemble the behaviour of cloth used by the tailor in the last step of the process.

Since we want to simulate the interaction of a garment with the animated avatar, we need to handle collisions. During simulation we check for each triangle if it intersects the body in its current pose using the signed distance field of the avatar. If we detect an intersection, parts of the garment are inside the avatar and we resolve them by moving the garment back to the outside. It would be easy to add cloth-cloth collisions as well, however, these collisions do not introduce significant internal forces and therefore hardly affect the results. For this reason we made the design decision to omit the costly cloth-cloth collision detection step.

\subsubsection{Per triangle stretch} \label{sec:stretch}

\begin{figure}[b]
	\centering
	\includegraphics[width=0.70\linewidth]{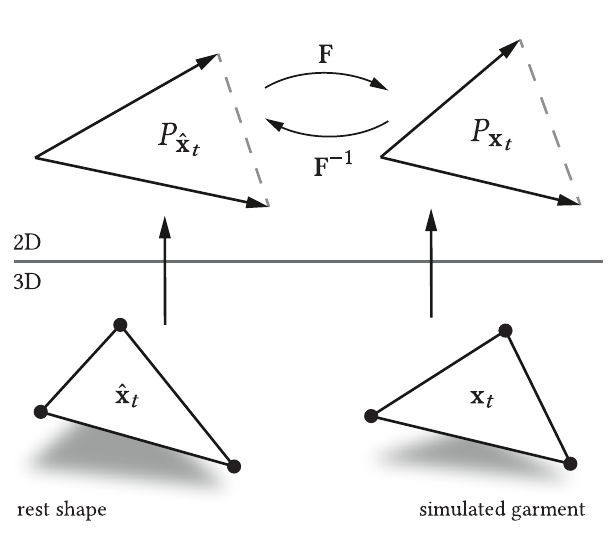}
	\caption{A 3D triangle, given by three points, can be rigidly mapped to a 2D triangle represented by two vectors. The deformation gradient relates the 2D triangles of the rest shape and the simulated garment.}
	\label{fig:deformation_gradient}
\end{figure}

Here we only focus on the stretch energy since it is the key to our garment adjustment step introduced in the following section. We refer to \cite{baraff1998large} for more details about shearing and bending terms. We consider a rest shape triangle $\hat{\mathbf x}_t$ and the corresponding simulation triangle $\mathbf x_t$. Since stretch is invariant to rigid transformations, we can consider a projection of these triangles to the 2D plane $P_{\mathbf x_t}$ and $P_{\hat{\mathbf x}_t}$. The \emph{deformation gradient} $\mathbf F \in \mathds R^{2\times 2}$ maps between these 2D triangles
\begin{align}\label{eq:defgrad}
     \mathbf F P_{\hat{\mathbf x}_t} = P_{\mathbf x_t}.
\end{align}
If $\mathbf F$ is a pure rotation, no stretch is exerted and we quantify stretch by measuring how much $\mathbf F$ differs from a rotation. To this end we consider the singular values of $\mathbf F$ (see appendix \ref{app:stretch}).

\subsubsection{Triangle adaption} \label{sec:rest_shape2}
 
An avatar raising its arms, for example, will introduce stress on the shirt that can not be reduced by the cloth simply moving with the body. The cloth simulation is based on the rest shape, but only modifies the simulation mesh and cannot counteract stretch due to larger movements. To optimize for dynamic stretch we want to optimize the rest shape mesh such that the garment allows for that movement during simulation and stretch is limited to small values, within the interval $\left[0, 1 + \delta\right]$. In that sense both the cloth simulation and the shape adaption are complementary problems. We can use this observation to answer the question: How do we have to modify the rest shape triangles such that that stretch is limited, assuming that we keep the simulation triangles fixed? By inverting $\mathbf F$, we can express the rest shape triangle with the simulation triangle
\begin{align}\label{eq:defgrad2}
      P_{\hat{\mathbf x}_t} = \mathbf F^{-1} P_{\mathbf x_t}.
\end{align}
The key idea now is to build a new deformation gradient $\bar{\mathbf F}$ that has two properties: (1) it's stretch is bound within $\left[0, 1 + \delta\right]$ and (2) it is as close as possible to the original deformation gradient $\mathbf F$. The desired matrix $\bar{\mathbf F}$ can be found using singular value decomposition and we give details on the construction in appendix \ref{app:stretch}. Now we obtain our updated rest shape triangle by computing
\begin{align}
     \bar{P}_{\hat{\mathbf x}_t} = \bar{\mathbf  F}^{-1} P_{\mathbf x_t}.
\end{align}
This new triangle has the desired properties, however, we are modifying single triangles without ensuring that they form a consistent mesh. We reconstruct a valid rest shape by stitching all triangles together using as-rigid-as-possible surface modeling \cite{ARAP_modeling:2007}. We provide more details in appendix \ref{app:arap}.

\begin{figure}[t]
	\centering
	\begin{overpic}[width=\linewidth]{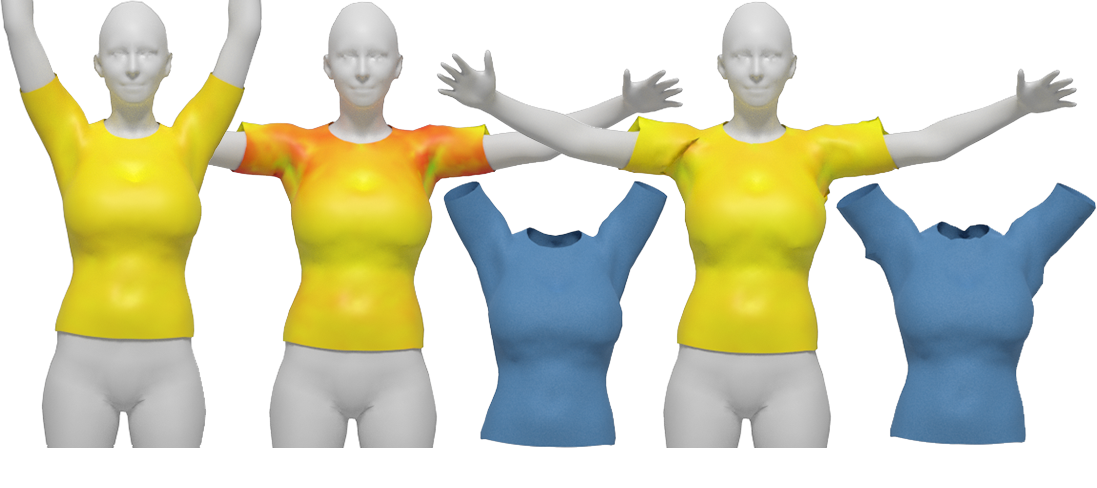}
	\put(9, 0){\small (a)}
	\put(30, 0){\small (b) }
	\put(48, 0){\small (c) }
	\put(66, 0){\small (d) }
	\put(86, 0){\small (e)}
	\end{overpic}
	\caption{A shirt (c) with a rest shape that was designed for raised arms (a) and therefore does not display any significant stretch in that pose, stretches once the arms are lowered (b). Red areas signify high stretch, as seen on top of the shoulders, whereas green areas signify compression, as seen in the armpits; yellow denotes no stretch. Once we deform the rest shape according to  \secref{sec:rest_shape2}~(e), the stretch is reduced significantly, as seen in~(d). In this example, we pinned the neckline to the avatar with the \textit{Pinning tool}}
	\label{fig:stretch}
\end{figure}
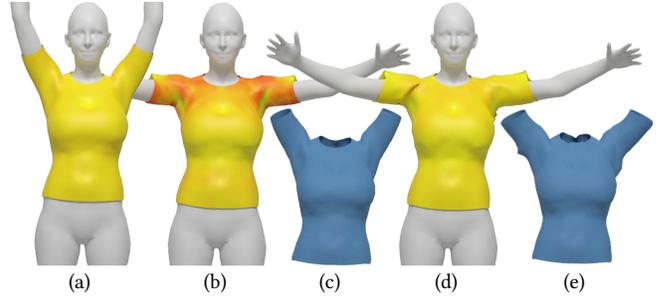

With this rest shape adaption strategy our algorithm for limiting dynamic stretch is straight forward. We simulate the garment on the animated avatar cycling through a fixed set of poses. As a result we obtain a garment that accommodates all poses that have been selected with limited stretch. Figure \ref{fig:stretch} shows a simple example of how stretch appears in a garment when the avatar lowers their arms and how it is reduced by adapting the rest shape.

\subsection{Garment fabrication}
\label{sec:fabrication}

After the final rest shape of the garment is computed, we use it as a basis for further production. In principle, users can choose any fabrication method that suits their needs, such as creating seamless knitting patterns using the method of Wu et al. \shortcite{wu2019knittable}. In this paper, we opted to create all example garments by generating sewing patterns through variational surface cutting \cite{sharp2018variational}. This method parameterizes surfaces over flat domains by directly optimizing the distortion induced by cutting and flattening. We weigh the proposed Hencky energy against the cut lengths to balance the number of sewing pattern pieces versus the introduced distortion. 
\replace{We initialize the pattern through normal clustering, choose a length normalization weight of $5$ and Hencky distortion weight of $10$ and take $500$ steps.}{}
As was shown in \cite{sharp2018variational}, even short cuts can yield sewing pattern pieces that nicely approximate the rest shape. 
\rev{Since our rest shape already accounts for the draping effect, we do not need to consider draping when creating a 2D sewing pattern, unlike previous work.}
\change{Instead of variational surface cutting, concurrent work \cite{pietroni2022computational} on creating sewing patterns from 3D shapes could be used as well.}
\change{A margin needs to be left when physically sewing the patterns. This can either be done easily by the seamstress (which we opted for), or automatically as described in \cite{Igarashi:SeamAllowance:2008}.}
We found that the flattening introduces only a very small amount of distortion across all our examples. In \figref{fig:results_dress} we show an optimized and simulated dress (b) along with the same dress (f) generated from stitching the individual sewing pattern pieces (e) back together.


\section{Results}
\label{sec:results}

Our implementation is based on several existing libraries: \texttt{libigl} \cite{libigl}, \texttt{PMP} \cite{pmp-library}, \texttt{OpenMP} \cite{dagum1998openmp} and \texttt{Cholmod} \cite{chen2008algorithm}. For Figures \ref{fig:tools}, \ref{fig:offset} and \ref{fig:fixed_boundaries} we use avatars from the FAUST dataset \cite{bogo2014faust}. All other avatars have been scanned by us. We use a computer with a 12-core \unit[2.7]{GHz} CPU and \unit[64]{GB} memory. Our source code is available on GitHub \url{https://github.com/katjawolff/custom_fit_garments}.

\begin{figure}[t]
	\centering
	\begin{overpic}[width=\linewidth]{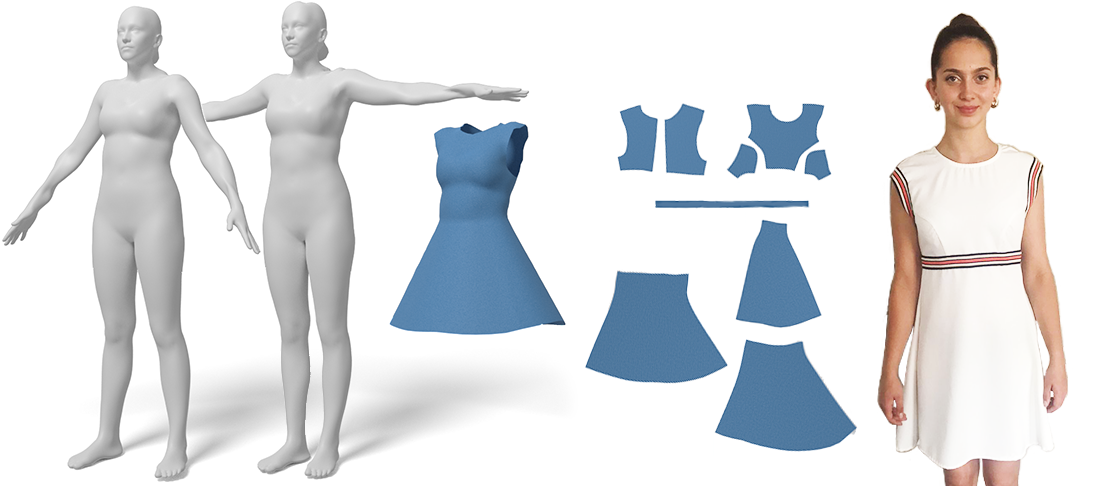}
	\put(18, 41){\small (a)}
	\put(42, 41){\small (b)}
	\put(64, 41){\small (c)}
	\put(93, 41){\small (d)}
	\end{overpic}
	\caption{A dress created from two poses (a). We show the final rest shape (b), the sewing pattern (c) and the physical garment (d).}
	\label{fig:results_extragirl}
\end{figure}
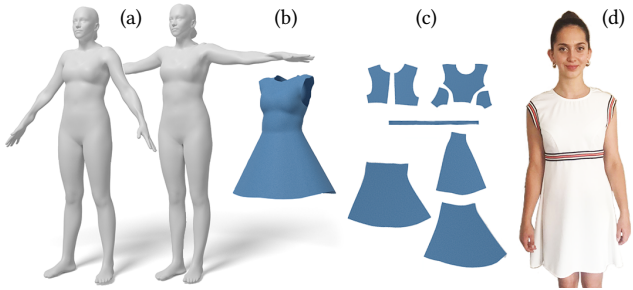

The dress shown in \figref{fig:results_extragirl} takes approximately 2.8 minutes to adjust to the two additional poses. Computation times vary mainly due to the varying number of poses and garment resolution.
In Table~\ref{table:performance} we report average frame rates for all modeling sessions presented in this paper. Our algorithm is implemented by augmenting a traditional Newton-based simulation framework, which is responsible for the bulk of computation time. Running the simulation without adapting the garment results in the performance reported under ``fps (sim)''; the full algorithm runs at a frame rate of ``fps (full)''. \replace{Consequently, any method accelerating the chosen type of cloth simulation technique immediately benefits our algorithm. The frame rate of our interactive application largely depends on the garment mesh resolution reported as the number of vertices.}{}

\begin{table}[b]
\centering
\caption{Performance measurements for all editing sessions. We report the number of vertices of the garment and average frames per second (fps), both for the full algorithm (fps (full)) and the simulation part only (fps (sim)).}
{\small
 \begin{tabular}{l c l l} 
 \toprule
Figure & \#vertices & fps (full) & fps (sim)\\
 \midrule
Fig.\ \ref{fig:results_sofia}        & 1103 & 7.82 & 8.29\\
Fig.\ \ref{fig:results_extraguy}     & 3926 & 1.70 & 1.74\\
Fig.\ \ref{fig:results_extragirl}    & 4270 & 1.76 & 1.81\\
Fig.\ \ref{fig:results_shirt_tpose}  & 3032 & 3.75 & 4.00\\
Fig.\ \ref{fig:results_shirt}        & 3032 & 3.69 & 3.94\\
Fig.\ \ref{fig:results_dress}        & 1484 & 5.73 & 6.00\\
Fig.\ \ref{fig:results_jumpsuit}     & 3597 & 3.32 & 3.54\\

 \bottomrule
 \end{tabular}
}
 \label{table:performance}
\end{table}

For all examples shown in this paper we use the same cloth parameters, chosen to simulate textiles that practically do not stretch in order to highlight the capabilities of our method, as they require the largest adjustments to the rest shapes. However, our method works with any kind of textile, if the cloth simulation parameters are chosen accordingly. 
We use the stiffness constants $k_\text{stretch} = 800$, $k_\text{shear} = 200$ and $k_\text{bend} = 10^{-6}$ and the damping stiffness constants $k_\text{d,stretch} = 100$, $k_\text{d,shear} = 1$ and $k_\text{d,bend} = 10^{-5}$ for the stretch, shear and bend constraints, respectively (for details we refer to \cite{baraff1998large}). We advance the simulation with a time step of $h = 0.0025$ and adjust the rest shape every 8 time steps with a stretch threshold of $\delta = 0.1$. We move through $60$ interpolation steps between two poses. We empirically found these parameters to consistently yield good results\replace{, however, any other set of parameters might be chosen in order to achieve a different trade off between accuracy and performance}{}.

\begin{figure*}[ht!]
	\centering
	\begin{overpic}[width=0.94\linewidth]{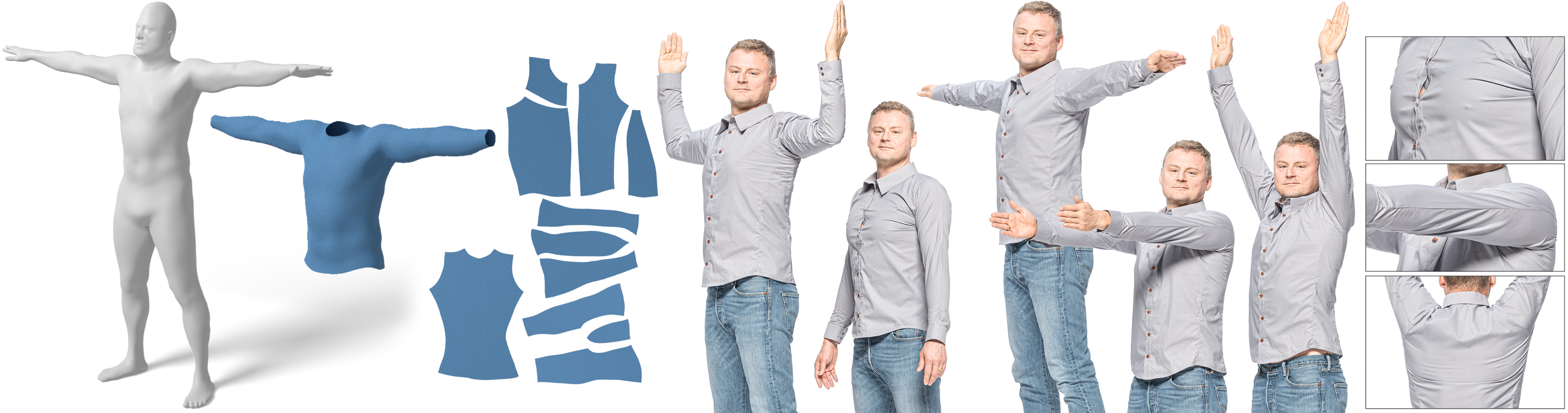}
	\put(13, 25){\small (a)}
	\put(22, 25){\small (b)}
	\put(36, 25){\small (c)}
	\put(70, 25){\small (d)}
	\put(92.5, 25){\small (e)}
	\end{overpic}
	\captionof{figure}{Starting from a single T-pose (a), we create the shape of a shirt (b) and its sewing pattern (c). This shirt fits perfectly in the T-pose, but creates uncomfortable stretch in other poses (d). Especially when the arms are held at the sides, visible stress is produced on the buttons of the shirt. This is also noticeable for arms outstretched to the front and up (e).}
	\label{fig:results_shirt_tpose}
\vspace{1em}
	\centering
	\begin{overpic}[width=0.95\linewidth]{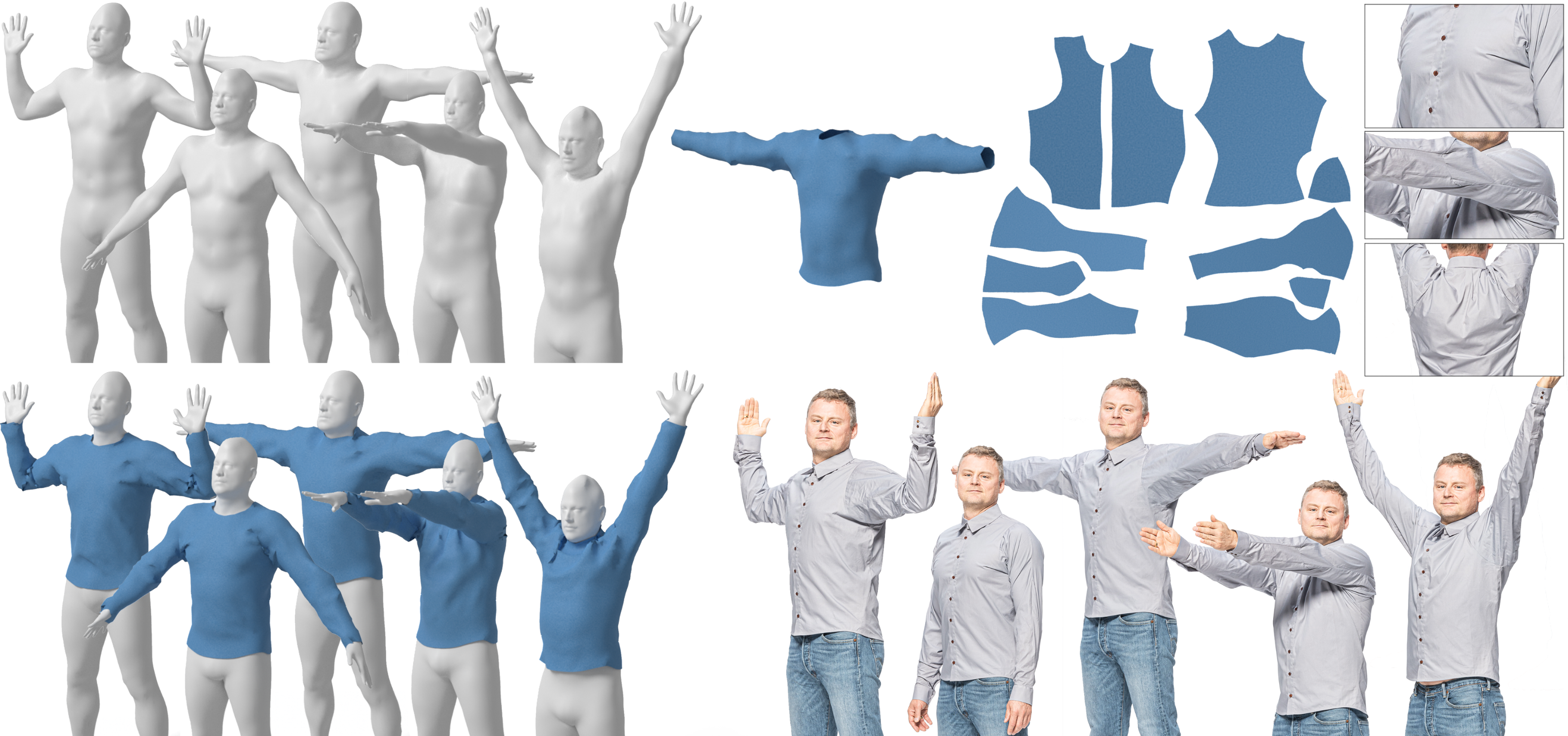}
	\put(36, 45){\small (a)}
	\put(52, 45){\small (b)}
	\put(74, 45){\small (c)}
	\put(85, 45){\small (d)}
	\put(41, 0){\small (e)}
	\put(47.5, 0){\small (f)}
	\end{overpic}
	\captionof{figure}{Starting from the same shirt design for the T-pose as in {\protect \figref{fig:results_shirt_tpose}}, we now move through five different poses (a) to update the rest shape of the garment (b) and create a sewing pattern for fabrication (c). We show the final digital garment on all five poses (e), as well as the sewn shirt (f). The shirt accommodates all five poses, and no uncomfortable stretch occurs (d). }
	\label{fig:results_shirt}
\end{figure*}

We created several  garments and computed the sewing patterns as detailed in \secref{sec:fabrication}. All are sewn by professional tailors. We list the design parameters for the individual garments in Table \ref{table:1}. 

\paragraph*{Men's shirt.}
To highlight the benefit of pose adaptive garment design we created two men's shirts for the same person. The first one, shown in \figref{fig:results_shirt_tpose}, is created from a single T-pose, while the second one is based on five different poses shown in \figref{fig:results_shirt}, starting from the T-pose. For both shirts we use the \textit{Boundary tool} to create the shape of the garment and add an offset of \unit[1.5]{cm} with the \textit{Comfort tool}. 
We pin the collar and cuffs with the \textit{Pinning tool} to ensure that the fabric of the sleeves is stretched sufficiently instead of being pulled back when moving the arms. Finally, we pre-define the shoulder seams and button border typical for shirts with the \textit{Seam tool}. The sewing pattern is then created from this pre-cut mesh \replace{(here we use a Hencky distortion weight of $3$ to create fewer seams)}{}. A professional tailor added the collar, cuffs and button border. A comparison of both shirts (\figref{fig:results_shirt_tpose} (d) and \figref{fig:results_shirt} (e)) shows that the shirt created solely from the T-pose fits in this specific pose and stretches uncomfortably in other poses, close to tearing. The second shirt allows a wider range of motions. \figref{fig:compare_shirts} shows additional photographs of the second shirt and compares to an off-the-shelf, standard shirt for the same person. Both shirts take approximately 7 hours to fabricate, from cutting to adding buttons.
According to the tailors we worked with, the production time for garments depends mostly on the number of pieces and the length of the seams. Therefore they estimate that the production time for our shirt and a standard one would not differ significantly.

\begin{figure*}[t]
	\centering
	\begin{overpic}[width=0.97\linewidth]{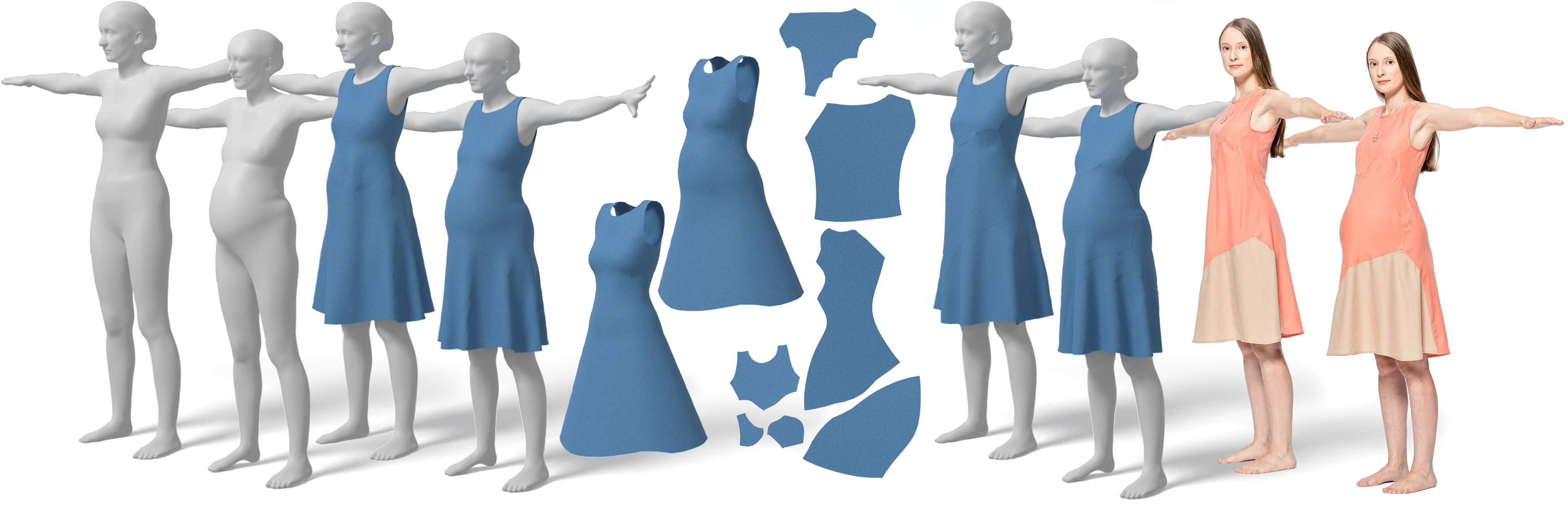}
	\put(11,  0){\small (a)}
	\put(27  ,0){\small (b)}
	\put(39.5,0){\small (c)}
	\put(46,  0){\small (d)}
	\put(53,0){\small (e)}
	\put(66,0){\small (f)}
	\put(84.5,0){\small (g)}
	\end{overpic}
	
	\caption{Our method can  be used to create garments that fit during and after pregnancy. Instead of different poses, we start from different scanned body shapes (a). We design the initial dress shape on the slim pose (c), compute its adjustment to both poses to get the final shape (d) and create a sewing pattern (e). Comparison of the draped final rest shape on both poses (b) with the sewn and draped pattern (f) reveals only negligible difference. The sewn garment (g) can be worn during and after pregnancy. }
	\label{fig:results_dress}
\end{figure*}

\paragraph*{Maternity wear.}
Our method can also be used to make garments for special body shapes. We create a dress that fits during \replace{}{and after} pregnancy \replace{and afterwards}{} (\figref{fig:results_dress})\replace{, starting from the slim shape}{}. After the rest shape adjustment from the slim shape to the pregnant one, we additionally adjust the rest shape slightly by increasing the cloth area using the \textit{Paint tool} below the belly for a smother transition. In \figref{fig:results_dress} (b) and (e) we compare the garment shape before generating the sewing pattern and after cutting, flattening and re-simulating in 3D. The 3D garment shape does not change visibly.

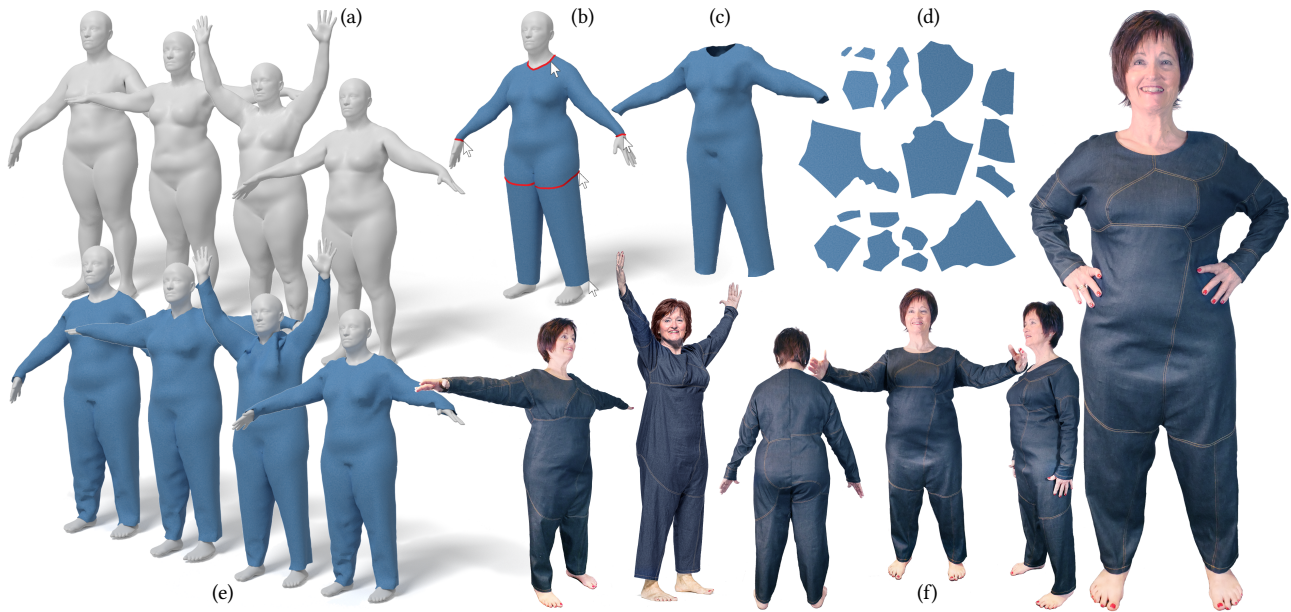
\begin{figure*}[t]
	\centering
		\begin{overpic}[width=0.95\linewidth]{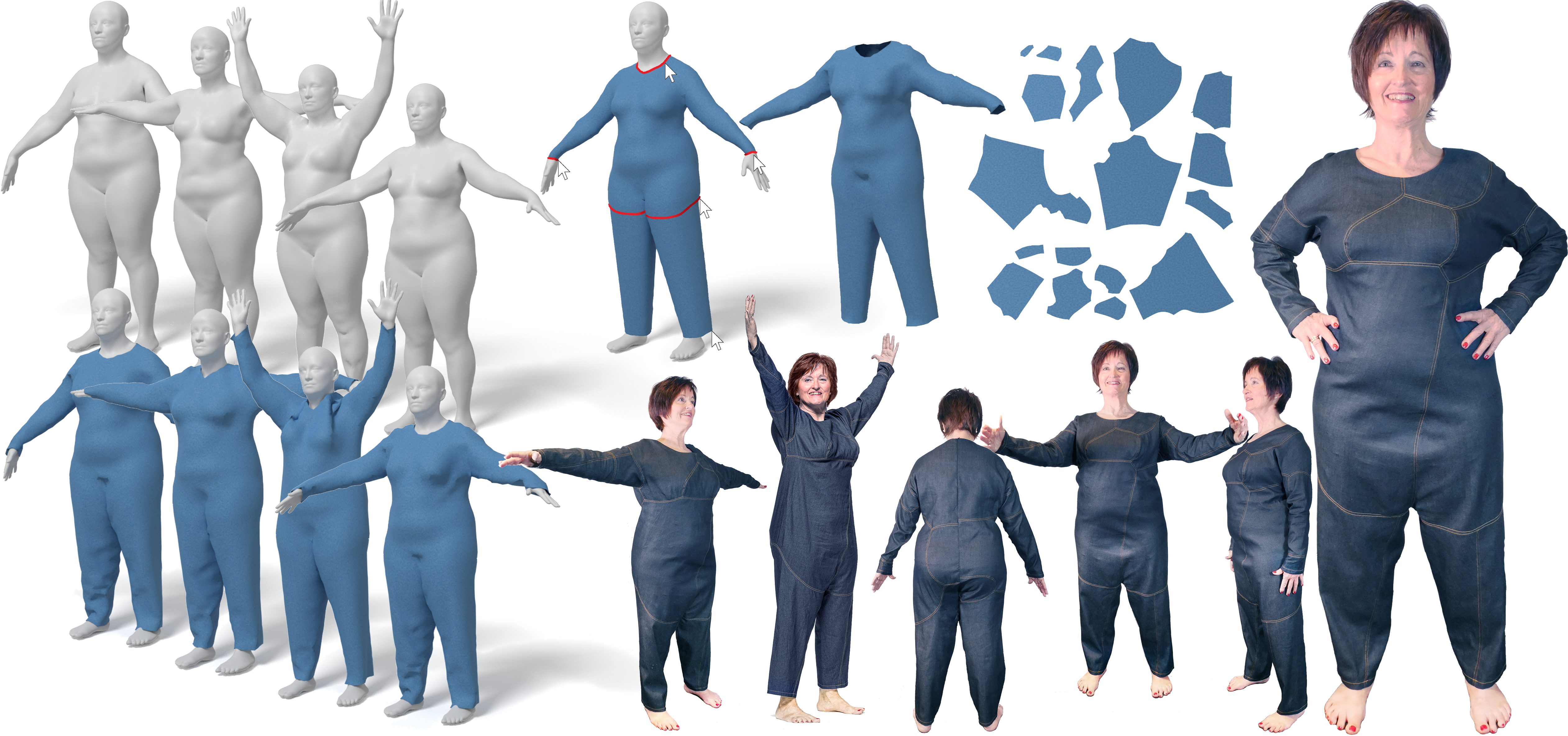}
    	\put(26, 46){\small (a)}
    	\put(44 ,46){\small (b)}
    	\put(54.8,46){\small (c)}
    	\put(71, 46){\small (d)}
    	\put(16,1){\small (e)}
    	\put(71,1){\small (f)}
	\end{overpic}
	\caption{The jumpsuit constitutes a challenging full-body example, where we use four poses as input (a). Note that in the first pose the arms are lowered halfway at the sides, whereas in the last pose the arms are held halfway to the front. We use our toolset to create an initial garment (b) and compute the adjustment of the rest shape to all four poses (c), which we drape on all four avatar poses (e) and use to create a sewing pattern (d). The physical jumpsuit fits all poses comfortably (f).}
	\label{fig:results_jumpsuit}
\end{figure*}

\paragraph*{Jumpsuit.}
By creating a jumpsuit, we show that our method can handle complex cases of larger garments that cover the whole body (\figref{fig:results_jumpsuit}). We again use the \textit{Boundary tool} to define the upper part of the jumpsuit, and the \textit{Extension tool} to create the legs. Starting from the A-pose with half-raised arms at the sides, we move through 4 poses, while pinning the collar and cuffs with the \textit{Pinning tool} and adding an offset of \unit[1]{cm} with the \textit{Comfort tool}. Using the \textit{Seam tool}, we create a straight seam in the back for a zipper. Fabricating the denim jumpsuit takes 14 hours, where half of the time is used to create the elaborate seaming. 

\paragraph*{Inclusive garment design.}
The strength of our method is showcased by creating garments for people who fall far outside the standard sizes available in stores, see \figref{fig:results_sofia}. In this example, imperfections in the registered avatars for the different poses do appear, especially around the hands and feet, likely due to the SMPL model not being trained on such body shapes. Still, our method is robust to small scan errors, and we can still design a custom-fit dress.

\begin{figure}[t]
	\centering
	\begin{overpic}[width=\linewidth]{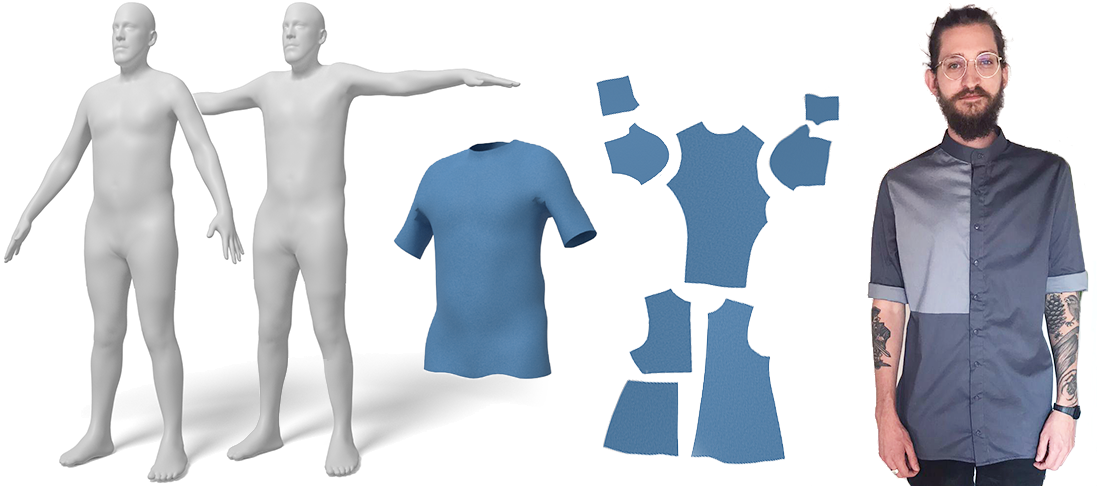}
	\put(18, 41){\small (a)}
	\put(43, 41){\small (b)}
	\put(64, 41){\small (c)}
	\put(94, 41){\small (d)}
	\end{overpic}
	\caption{\label{fig:results_extraguy}A shirt created from two poses (a). We show the final rest shape (b), the sewing pattern (c) and the physical garment (d).}
	
\end{figure}

\paragraph*{Additional results.}
We show two additional simple results, created from just two poses: a shirt (\figref{fig:results_extraguy}) and a white dress (\figref{fig:results_extragirl}), to highlight the variety of possible garments. The shirt in \figref{fig:results_extraguy} is designed in the same way as the previous shirt from \figref{fig:results_shirt}. The white dress is created similarly to the pregnancy dress from \figref{fig:results_dress}, but we use the \textit{Seam tool} to cut a strip of cloth that is replaced by a red band. By calculating the underlying garment shape, but allowing the professional tailor to add small details like buttons, trims and collars, a diverse set of garments can be created. 

\subsection{User feedback} \label{sec:study}
Our method was evaluated by producing garments for six people of varying body shapes. The participants were between 19 and 63 years old with their heights ranging from \unit[125]{cm} to \unit[190]{cm}. Five participants filled out a questionnaire about their experience with our produced garments and fashion in general. All questions were answered on a 7-point scale from \emph{1: I strongly disagree} to \emph{7: I strongly agree}.  
On the question ``\emph{Do you regularly feel restricted due to tight clothing?}'', participants where slightly positive (M=4.8, SD=1.3), indicating that the issue arises occasionally. All participants had experience with online shopping of textiles and responded mostly positive (M=5.4, SD=1.52) to the question ``\emph{Do you often return online ordered clothes because they do not fit as expected?}''. Asked about the fit of their customized garment, participants responded positively (M=6.4, SD=0.89). Comfort was  consistently rated even higher (M=6.6, SD=0.55). The final garment was unanimously perceived as being consistent with the digital design (M=7, SD=0). The question ``\emph{How do you rate freedom of movement while wearing the garment?}'', was answered affirmatively by the participants (M=6.6, SD=0.55). 
Asked about their overall experience, one participant responded that they  \emph{generally have problems finding fitting clothes that are also fashionable and unique} and hopes that customized clothing will \emph{help her save time on shopping}. Another participant mentioned that the garment was particularly comfortable to wear because the sewing pattern was non-symmetric and that they \emph{learned about the fact that they had a slightly asymmetric chest by seeing the sewing pattern}.

\subsection{Discussion}
The choice of the initial pose plays a vital role in the final garment shape, as the garment exhibits the least amount of folds in this pose. The further the avatar moves away from that pose, the more folds can be expected. This can be seen when comparing the shirt in \figref{fig:results_shirt}, which is designed for the T-pose, and the shirt in \figref{fig:results_extraguy}, which is designed for the A-pose. There is a tradeoff between fit and movement range that has to be considered when using our system. The more poses the garment needs to fit tightly in, the less the garment can fit each individual pose. For example, traditionally designed shirts usually do not accommodate a pose with arms raised high above the head. \rev{The shirt in \figref{fig:results_shirt} accommodates poses that would not be possible in a standard shirt, resulting in more surface area around the elbows and shoulders. By discarding these extra poses, we can achieve a tighter fit with a limited movement range, which most people are already accustomed to (\figref{fig:results_extraguy}).}

\section{Conclusion}

We presented a method to design and optimize the shape of a garment for a range of 3D scanned poses, such that the garment fits comfortably but tightly in all poses. We demonstrated the capabilities of our method by sewing different garments for a range of individual body shapes.

\paragraph*{Limitations.}

Since the input of our method is a number of registered 3D scans, we are limited by the current state of the art in human body model capture. This becomes apparent when we scan people with proportions strongly deviating from the average, as can be seen in \figref{fig:results_sofia}, where difficulties in registering the scans result in elongated hands and misplaced heels arise. In theory, we could design garments for people with missing extremities, but current methods based on SMPL \cite{SMPL:2015} would hallucinate these extremities, requiring manual fixing of the body meshes. \replace{The same registration process also creates self-intersections in the avatars, leading to problems in cloth simulation and precluding the usage of self-intersecting poses as the initial poses for designing a garment.}{}

\paragraph*{Future work.}
Currently, the final shape of the garment depends on the sequence of poses and especially on the initial one. Though this gives designers the option to choose a main pose in which the garment is worn, creating a garment independent from the simulation sequence might consider the poses more equally. 
Furthermore, we initiate the rest shape of the garment from the body shape of the initial pose and do not take the draping effect into account in this first step. Generating the initial rest shape such that it physically deforms into a tight fit could slightly improve our method in larger convex areas, like under breasts and the bottom.

Our method only allows for the body to influence the garment shape, but not vice versa. In the future, we would like to incorporate the physical simulation of body tissue to simulate the influence of the garment on the body, 
\change{by using the dynamic body model Dyna \cite{Dyna:SIGGRAPH:2015} or the STAR model \cite{osman2020star}.}
\replace{Especially in tight areas, body tissue can be  visibly displaced by the garment.
Incorporating recent work in body modeling, like the dynamic body model Dyna \cite{Dyna:SIGGRAPH:2015} or the STAR model \cite{osman2020star}, would be a first step in that direction.}{} 
We would also like to improve the simulation by incorporating cloth parameters measured from real textiles.
Incorporating friction into the simulation would further improve the design of pants or skirts to prevent rubbing against the skin and sliding.

\paragraph*{Acknowledgements}
We would like to thank all our models and tailors that supported our project. We thank Etienne Vouga for insightful discussions. This work was partially supported by the Personalized Health and Related Technologies (PHRT) SwissHeart grant and the European Research Council (ERC) under the European Union’s Horizon 2020 research and innovation programme (ERC Consolidator Grant, agreement
No. 101003104, MYCLOTH).

\begin{table*}[t]
\centering
\caption{Statistics for all garments.\vspace{-7pt}}
{\small
 \begin{tabular}{l l r r l l r l r r r} 
 \toprule
  & Figure & faces & poses & initial pose & tools & offset & textile & pieces & sewing time & model height\\
 \midrule
 Shirt T-pose       & \ref{fig:results_shirt_tpose}  & \unit[6]{k} & 1 & T-pose & 1,4,6 & \unit[1.5]{cm} & cotton poplin & 11 & \unit[7]{h} & \unit[190]{cm}\\
 Shirt              & \ref{fig:compare_shirts}, \ref{fig:results_shirt} & \unit[6]{k} & 5 & T-pose & 1,4,5,6 & \unit[1.5]{cm} & cotton poplin & 10 & \unit[6.5]{h} & \unit[190]{cm}\\
 Jumpsuit           & \ref{fig:results_jumpsuit}   & \unit[7]{k} & 4 & A-pose & 1,2,4,5,6 & \unit[1.0]{cm} &  denim & 17 & \unit[14]{h} & \unit[163]{cm}\\
 Pregnancy dress    & \ref{fig:results_dress}   & \unit[4]{k} & 2 & pre-pregnancy & 1,2,3,4 & \unit[0.8]{cm} &  lyocell & 7 & \unit[3]{h} & \unit[172]{cm}\\
 Yellow dress       & \ref{fig:results_sofia} & \unit[5]{k} & 5 & T-pose & 1,2,4,5,6 & \unit[0.5]{cm} &  lyocell & 13 & \unit[2.75]{h} & \unit[125]{cm}\\
 White dress        & \ref{fig:results_extragirl}  & \unit[9]{k}  & 2 & A-pose & 1,2,4,6 & \unit[0.5]{cm} &  polyester crepe & 9 & \unit[3]{h} & \unit[177]{cm}\\
 Blue Shirt         & \ref{fig:results_extraguy}  & \unit[8]{k}  & 2 & A-pose & 1,4,5,6 & \unit[1.0]{cm} & cotton poplin & 8 & \unit[4.5]{h} & \unit[188]{cm}\\
 \bottomrule
 \end{tabular}
 }
 \label{table:1}
 \vspace{-3pt}
\end{table*}

\bibliographystyle{ACM-Reference-Format}
\bibliography{paper}


\begin{thebibliography}{64}


\ifx \showCODEN    \undefined \def \showCODEN     #1{\unskip}     \fi
\ifx \showDOI      \undefined \def \showDOI       #1{#1}\fi
\ifx \showISBNx    \undefined \def \showISBNx     #1{\unskip}     \fi
\ifx \showISBNxiii \undefined \def \showISBNxiii  #1{\unskip}     \fi
\ifx \showISSN     \undefined \def \showISSN      #1{\unskip}     \fi
\ifx \showLCCN     \undefined \def \showLCCN      #1{\unskip}     \fi
\ifx \shownote     \undefined \def \shownote      #1{#1}          \fi
\ifx \showarticletitle \undefined \def \showarticletitle #1{#1}   \fi
\ifx \showURL      \undefined \def \showURL       {\relax}        \fi
\providecommand\bibfield[2]{#2}
\providecommand\bibinfo[2]{#2}
\providecommand\natexlab[1]{#1}
\providecommand\showeprint[2][]{arXiv:#2}

\bibitem[\protect\citeauthoryear{Baraff and Witkin}{Baraff and Witkin}{1998}]%
        {baraff1998large}
\bibfield{author}{\bibinfo{person}{David Baraff} {and} \bibinfo{person}{Andrew
  Witkin}.} \bibinfo{year}{1998}\natexlab{}.
\newblock \showarticletitle{Large steps in cloth simulation}. In
  \bibinfo{booktitle}{\emph{Proc.\ SIGGRAPH}}. \bibinfo{pages}{43--54}.
\newblock


\bibitem[\protect\citeauthoryear{Bartle, Sheffer, Kim, Kaufman, Vining, and
  Berthouzoz}{Bartle et~al\mbox{.}}{2016}]%
        {Bartle:PhysicsPatternAdj:2016}
\bibfield{author}{\bibinfo{person}{Aric Bartle}, \bibinfo{person}{Alla
  Sheffer}, \bibinfo{person}{Vladimir~G Kim}, \bibinfo{person}{Danny~M
  Kaufman}, \bibinfo{person}{Nicholas Vining}, {and} \bibinfo{person}{Floraine
  Berthouzoz}.} \bibinfo{year}{2016}\natexlab{}.
\newblock \showarticletitle{Physics-driven pattern adjustment for direct 3D
  garment editing.}
\newblock \bibinfo{journal}{\emph{ACM Trans.\ Graph.}} \bibinfo{volume}{35},
  \bibinfo{number}{4} (\bibinfo{year}{2016}).
\newblock


\bibitem[\protect\citeauthoryear{Bergou, Wardetzky, Harmon, Zorin, and
  Grinspun}{Bergou et~al\mbox{.}}{2006}]%
        {bergou2006quadratic}
\bibfield{author}{\bibinfo{person}{Miklos Bergou}, \bibinfo{person}{Max
  Wardetzky}, \bibinfo{person}{David Harmon}, \bibinfo{person}{Denis Zorin},
  {and} \bibinfo{person}{Eitan Grinspun}.} \bibinfo{year}{2006}\natexlab{}.
\newblock \showarticletitle{A quadratic bending model for inextensible
  surfaces}. In \bibinfo{booktitle}{\emph{Proc.\ SGP}}.
  \bibinfo{pages}{227--230}.
\newblock


\bibitem[\protect\citeauthoryear{Berthouzoz, Garg, Kaufman, Grinspun, and
  Agrawala}{Berthouzoz et~al\mbox{.}}{2013}]%
        {Berthouzoz:ParsingPatterns:2013}
\bibfield{author}{\bibinfo{person}{Floraine Berthouzoz}, \bibinfo{person}{Akash
  Garg}, \bibinfo{person}{Danny~M Kaufman}, \bibinfo{person}{Eitan Grinspun},
  {and} \bibinfo{person}{Maneesh Agrawala}.} \bibinfo{year}{2013}\natexlab{}.
\newblock \showarticletitle{Parsing sewing patterns into 3D garments}.
\newblock \bibinfo{journal}{\emph{ACM Trans.\ Graph.}} \bibinfo{volume}{32},
  \bibinfo{number}{4} (\bibinfo{year}{2013}).
\newblock


\bibitem[\protect\citeauthoryear{Bick, Halsey, and Ekenga}{Bick
  et~al\mbox{.}}{2018}]%
        {bick2018global}
\bibfield{author}{\bibinfo{person}{Rachel Bick}, \bibinfo{person}{Erika
  Halsey}, {and} \bibinfo{person}{Christine~C Ekenga}.}
  \bibinfo{year}{2018}\natexlab{}.
\newblock \showarticletitle{The global environmental injustice of fast
  fashion}.
\newblock \bibinfo{journal}{\emph{Environmental Health}} \bibinfo{volume}{17},
  \bibinfo{number}{1} (\bibinfo{year}{2018}), \bibinfo{pages}{92}.
\newblock


\bibitem[\protect\citeauthoryear{Bogo, Romero, Loper, and Black}{Bogo
  et~al\mbox{.}}{2014}]%
        {bogo2014faust}
\bibfield{author}{\bibinfo{person}{Federica Bogo}, \bibinfo{person}{Javier
  Romero}, \bibinfo{person}{Matthew Loper}, {and} \bibinfo{person}{Michael~J
  Black}.} \bibinfo{year}{2014}\natexlab{}.
\newblock \showarticletitle{FAUST: Dataset and evaluation for 3D mesh
  registration}. In \bibinfo{booktitle}{\emph{Proceedings of the IEEE
  Conference on Computer Vision and Pattern Recognition}}.
  \bibinfo{pages}{3794--3801}.
\newblock


\bibitem[\protect\citeauthoryear{Bridson, Marino, and Fedkiw}{Bridson
  et~al\mbox{.}}{2005}]%
        {bridson2005simulation}
\bibfield{author}{\bibinfo{person}{Robert Bridson}, \bibinfo{person}{Sebastian
  Marino}, {and} \bibinfo{person}{Ronald Fedkiw}.}
  \bibinfo{year}{2005}\natexlab{}.
\newblock \showarticletitle{Simulation of clothing with folds and wrinkles}.
\newblock In \bibinfo{booktitle}{\emph{ACM SIGGRAPH 2005 Courses}}.
  \bibinfo{pages}{3--es}.
\newblock


\bibitem[\protect\citeauthoryear{Brouet, Sheffer, Boissieux, and Cani}{Brouet
  et~al\mbox{.}}{2012}]%
        {Brouet:DesignPresGarTrans:2012}
\bibfield{author}{\bibinfo{person}{Remi Brouet}, \bibinfo{person}{Alla
  Sheffer}, \bibinfo{person}{Laurence Boissieux}, {and}
  \bibinfo{person}{Marie-Paule Cani}.} \bibinfo{year}{2012}\natexlab{}.
\newblock \showarticletitle{Design preserving garment transfer}.
\newblock \bibinfo{journal}{\emph{ACM Trans.\ Graph.}} \bibinfo{volume}{31},
  \bibinfo{number}{4} (\bibinfo{year}{2012}).
\newblock


\bibitem[\protect\citeauthoryear{Chen, Davis, Hager, and Rajamanickam}{Chen
  et~al\mbox{.}}{2008}]%
        {chen2008algorithm}
\bibfield{author}{\bibinfo{person}{Yanqing Chen}, \bibinfo{person}{Timothy~A
  Davis}, \bibinfo{person}{William~W Hager}, {and}
  \bibinfo{person}{Sivasankaran Rajamanickam}.}
  \bibinfo{year}{2008}\natexlab{}.
\newblock \showarticletitle{Algorithm 887: CHOLMOD, supernodal sparse Cholesky
  factorization and update/downdate}.
\newblock \bibinfo{journal}{\emph{ACM Trans. Math. Software}}
  \bibinfo{volume}{35}, \bibinfo{number}{3} (\bibinfo{year}{2008}),
  \bibinfo{pages}{1--14}.
\newblock


\bibitem[\protect\citeauthoryear{Choi and Ko}{Choi and Ko}{2005}]%
        {choi2005stable}
\bibfield{author}{\bibinfo{person}{Kwang-Jin Choi} {and}
  \bibinfo{person}{Hyeong-Seok Ko}.} \bibinfo{year}{2005}\natexlab{}.
\newblock \showarticletitle{Stable but responsive cloth}.
\newblock In \bibinfo{booktitle}{\emph{ACM SIGGRAPH 2005 Courses}}.
  \bibinfo{pages}{1--es}.
\newblock


\bibitem[\protect\citeauthoryear{Cirio, Lopez-Moreno, Miraut, and Otaduy}{Cirio
  et~al\mbox{.}}{2014}]%
        {cirio2014yarn}
\bibfield{author}{\bibinfo{person}{Gabriel Cirio}, \bibinfo{person}{Jorge
  Lopez-Moreno}, \bibinfo{person}{David Miraut}, {and}
  \bibinfo{person}{Miguel~A Otaduy}.} \bibinfo{year}{2014}\natexlab{}.
\newblock \showarticletitle{Yarn-level simulation of woven cloth}.
\newblock \bibinfo{journal}{\emph{ACM Trans.\ Graph.}} \bibinfo{volume}{33},
  \bibinfo{number}{6} (\bibinfo{year}{2014}), \bibinfo{pages}{1--11}.
\newblock


\bibitem[\protect\citeauthoryear{CLO}{CLO}{2020}]%
        {CLO3D}
\bibfield{author}{\bibinfo{person}{CLO}.} \bibinfo{year}{2020}\natexlab{}.
\newblock \bibinfo{title}{clo3d.com}.
\newblock \bibinfo{howpublished}{\url{https://www.clo3d.com}}.
\newblock


\bibitem[\protect\citeauthoryear{Cordier, Seo, and Magnenat-Thalmann}{Cordier
  et~al\mbox{.}}{2003}]%
        {cordier2003made}
\bibfield{author}{\bibinfo{person}{Frederic Cordier}, \bibinfo{person}{Hyewon
  Seo}, {and} \bibinfo{person}{Nadia Magnenat-Thalmann}.}
  \bibinfo{year}{2003}\natexlab{}.
\newblock \showarticletitle{Made-to-measure technologies for an online clothing
  store}.
\newblock \bibinfo{journal}{\emph{IEEE Computer graphics and applications}}
  \bibinfo{volume}{23}, \bibinfo{number}{1} (\bibinfo{year}{2003}),
  \bibinfo{pages}{38--48}.
\newblock


\bibitem[\protect\citeauthoryear{Dagum and Menon}{Dagum and Menon}{1998}]%
        {dagum1998openmp}
\bibfield{author}{\bibinfo{person}{Leonardo Dagum} {and}
  \bibinfo{person}{Ramesh Menon}.} \bibinfo{year}{1998}\natexlab{}.
\newblock \showarticletitle{OpenMP: an industry standard API for shared-memory
  programming}.
\newblock \bibinfo{journal}{\emph{Computational Science \& Engineering, IEEE}}
  \bibinfo{volume}{5}, \bibinfo{number}{1} (\bibinfo{year}{1998}),
  \bibinfo{pages}{46--55}.
\newblock


\bibitem[\protect\citeauthoryear{Decaudin, Julius, Wither, Boissieux, Sheffer,
  and Cani}{Decaudin et~al\mbox{.}}{2006}]%
        {Decaudin:VirtualGarments:2006}
\bibfield{author}{\bibinfo{person}{Philippe Decaudin}, \bibinfo{person}{Dan
  Julius}, \bibinfo{person}{Jamie Wither}, \bibinfo{person}{Laurence
  Boissieux}, \bibinfo{person}{Alla Sheffer}, {and}
  \bibinfo{person}{Marie-Paule Cani}.} \bibinfo{year}{2006}\natexlab{}.
\newblock \showarticletitle{Virtual garments: A fully geometric approach for
  clothing design}.
\newblock \bibinfo{journal}{\emph{Comput.\ Graph.\ Forum}}
  \bibinfo{volume}{25}, \bibinfo{number}{3} (\bibinfo{year}{2006}),
  \bibinfo{pages}{625--634}.
\newblock


\bibitem[\protect\citeauthoryear{Goldenthal, Harmon, Fattal, Bercovier, and
  Grinspun}{Goldenthal et~al\mbox{.}}{2007}]%
        {goldenthal2007efficient}
\bibfield{author}{\bibinfo{person}{Rony Goldenthal}, \bibinfo{person}{David
  Harmon}, \bibinfo{person}{Raanan Fattal}, \bibinfo{person}{Michel Bercovier},
  {and} \bibinfo{person}{Eitan Grinspun}.} \bibinfo{year}{2007}\natexlab{}.
\newblock \showarticletitle{Efficient Simulation of Inextensible Cloth}.
\newblock \bibinfo{journal}{\emph{ACM Trans. Graph.}} \bibinfo{volume}{26},
  \bibinfo{number}{3} (\bibinfo{date}{July} \bibinfo{year}{2007}),
  \bibinfo{pages}{49–es}.
\newblock
\showISSN{0730-0301}
\urldef\tempurl%
\url{https://doi.org/10.1145/1276377.1276438}
\showDOI{\tempurl}


\bibitem[\protect\citeauthoryear{Guan, Reiss, Hirshberg, Weiss, and Black}{Guan
  et~al\mbox{.}}{2012}]%
        {guan2012drape}
\bibfield{author}{\bibinfo{person}{Peng Guan}, \bibinfo{person}{Loretta Reiss},
  \bibinfo{person}{David~A Hirshberg}, \bibinfo{person}{Alexander Weiss}, {and}
  \bibinfo{person}{Michael~J Black}.} \bibinfo{year}{2012}\natexlab{}.
\newblock \showarticletitle{Drape: Dressing any person}.
\newblock \bibinfo{journal}{\emph{ACM Transactions on Graphics (TOG)}}
  \bibinfo{volume}{31}, \bibinfo{number}{4} (\bibinfo{year}{2012}),
  \bibinfo{pages}{1--10}.
\newblock


\bibitem[\protect\citeauthoryear{Igarashi, Igarashi, and Suzuki}{Igarashi
  et~al\mbox{.}}{2008}]%
        {Igarashi:SeamAllowance:2008}
\bibfield{author}{\bibinfo{person}{Yuki Igarashi}, \bibinfo{person}{Takeo
  Igarashi}, {and} \bibinfo{person}{Hiromasa Suzuki}.}
  \bibinfo{year}{2008}\natexlab{}.
\newblock \showarticletitle{Automatically adding seam allowance to cloth
  pattern}. In \bibinfo{booktitle}{\emph{ACM SIGGRAPH 2008 posters}}. ACM,
  \bibinfo{pages}{15}.
\newblock


\bibitem[\protect\citeauthoryear{Jacobson, Panozzo, et~al\mbox{.}}{Jacobson
  et~al\mbox{.}}{2018}]%
        {libigl}
\bibfield{author}{\bibinfo{person}{Alec Jacobson}, \bibinfo{person}{Daniele
  Panozzo}, {et~al\mbox{.}}} \bibinfo{year}{2018}\natexlab{}.
\newblock \bibinfo{title}{{libigl}: A simple {C++} geometry processing
  library}.
\newblock
\newblock
\newblock
\shownote{https://libigl.github.io/.}


\bibitem[\protect\citeauthoryear{Jung, Lee, Kim, Ryu, and Ko}{Jung
  et~al\mbox{.}}{2016}]%
        {jung2016modeling}
\bibfield{author}{\bibinfo{person}{Il-Hoe Jung}, \bibinfo{person}{Sang-Bin
  Lee}, \bibinfo{person}{Jong-Jun Kim}, \bibinfo{person}{Han-Na Ryu}, {and}
  \bibinfo{person}{Hyeong-Seok Ko}.} \bibinfo{year}{2016}\natexlab{}.
\newblock \showarticletitle{Modeling the non-elastic stretch deformation of
  cloth based on creep analysis}.
\newblock \bibinfo{journal}{\emph{Textile Research Journal}}
  \bibinfo{volume}{86}, \bibinfo{number}{3} (\bibinfo{year}{2016}),
  \bibinfo{pages}{245--255}.
\newblock


\bibitem[\protect\citeauthoryear{Kaldor, James, and Marschner}{Kaldor
  et~al\mbox{.}}{2010}]%
        {kaldor2010efficient}
\bibfield{author}{\bibinfo{person}{Jonathan~M Kaldor}, \bibinfo{person}{Doug~L
  James}, {and} \bibinfo{person}{Steve Marschner}.}
  \bibinfo{year}{2010}\natexlab{}.
\newblock \showarticletitle{Efficient yarn-based cloth with adaptive contact
  linearization}.
\newblock In \bibinfo{booktitle}{\emph{ACM SIGGRAPH 2010 papers}}.
  \bibinfo{pages}{1--10}.
\newblock


\bibitem[\protect\citeauthoryear{Keckeisen, Feurer, and Wacker}{Keckeisen
  et~al\mbox{.}}{2004}]%
        {Keckeisen:TailorTools:2004}
\bibfield{author}{\bibinfo{person}{Michael Keckeisen},
  \bibinfo{person}{Matthias Feurer}, {and} \bibinfo{person}{Markus Wacker}.}
  \bibinfo{year}{2004}\natexlab{}.
\newblock \showarticletitle{Tailor tools for interactive design of clothing in
  virtual environments}. In \bibinfo{booktitle}{\emph{Proceedings of the ACM
  symposium on Virtual reality software and technology}}. ACM,
  \bibinfo{pages}{182--185}.
\newblock


\bibitem[\protect\citeauthoryear{Kwok, Zhang, Wang, Liu, and Tang}{Kwok
  et~al\mbox{.}}{2016}]%
        {kwok:2016:styling}
\bibfield{author}{\bibinfo{person}{Tsz-Ho Kwok}, \bibinfo{person}{Yan-Qiu
  Zhang}, \bibinfo{person}{Charlie~CL Wang}, \bibinfo{person}{Yong-Jin Liu},
  {and} \bibinfo{person}{Kai Tang}.} \bibinfo{year}{2016}\natexlab{}.
\newblock \showarticletitle{Styling evolution for tight-fitting garments}.
\newblock \bibinfo{journal}{\emph{IEEE Trans.\ Vis.\ Comput.\ Graph.}}
  \bibinfo{volume}{22}, \bibinfo{number}{5} (\bibinfo{year}{2016}),
  \bibinfo{pages}{1580--1591}.
\newblock


\bibitem[\protect\citeauthoryear{Li, Sheffer, Grinspun, and Vining}{Li
  et~al\mbox{.}}{2018}]%
        {Li:2018:FoldSketch}
\bibfield{author}{\bibinfo{person}{Minchen Li}, \bibinfo{person}{Alla Sheffer},
  \bibinfo{person}{Eitan Grinspun}, {and} \bibinfo{person}{Nicholas Vining}.}
  \bibinfo{year}{2018}\natexlab{}.
\newblock \showarticletitle{FoldSketch: Enriching Garments with Physically
  Reproducible Folds}.
\newblock \bibinfo{journal}{\emph{ACM Trans.\ Graph.}} \bibinfo{volume}{37},
  \bibinfo{number}{4} (\bibinfo{year}{2018}).
\newblock
\urldef\tempurl%
\url{https://doi.org/10.1145/3197517.3201310}
\showDOI{\tempurl}


\bibitem[\protect\citeauthoryear{Liu, Bargteil, O'Brien, and Kavan}{Liu
  et~al\mbox{.}}{2013}]%
        {liu2013fast}
\bibfield{author}{\bibinfo{person}{Tiantian Liu}, \bibinfo{person}{Adam~W
  Bargteil}, \bibinfo{person}{James~F O'Brien}, {and} \bibinfo{person}{Ladislav
  Kavan}.} \bibinfo{year}{2013}\natexlab{}.
\newblock \showarticletitle{Fast simulation of mass-spring systems}.
\newblock \bibinfo{journal}{\emph{ACM Trans.\ Graph.}} \bibinfo{volume}{32},
  \bibinfo{number}{6} (\bibinfo{year}{2013}), \bibinfo{pages}{1--7}.
\newblock


\bibitem[\protect\citeauthoryear{Liu, Han, Zhang, Chen, Lai, Doubrovski,
  Whiting, and Wang}{Liu et~al\mbox{.}}{2021}]%
        {liu2021knitting}
\bibfield{author}{\bibinfo{person}{Zishun Liu}, \bibinfo{person}{Xingjian Han},
  \bibinfo{person}{Yuchen Zhang}, \bibinfo{person}{Xiangjia Chen},
  \bibinfo{person}{Yu-Kun Lai}, \bibinfo{person}{Eugeni~L Doubrovski},
  \bibinfo{person}{Emily Whiting}, {and} \bibinfo{person}{Charlie~CL Wang}.}
  \bibinfo{year}{2021}\natexlab{}.
\newblock \showarticletitle{Knitting 4D garments with elasticity controlled for
  body motion}.
\newblock \bibinfo{journal}{\emph{ACM Transactions on Graphics (TOG)}}
  \bibinfo{volume}{40}, \bibinfo{number}{4} (\bibinfo{year}{2021}),
  \bibinfo{pages}{1--16}.
\newblock


\bibitem[\protect\citeauthoryear{Loper, Mahmood, Romero, Pons-Moll, and
  Black}{Loper et~al\mbox{.}}{2015}]%
        {SMPL:2015}
\bibfield{author}{\bibinfo{person}{Matthew Loper}, \bibinfo{person}{Naureen
  Mahmood}, \bibinfo{person}{Javier Romero}, \bibinfo{person}{Gerard
  Pons-Moll}, {and} \bibinfo{person}{Michael~J. Black}.}
  \bibinfo{year}{2015}\natexlab{}.
\newblock \showarticletitle{{SMPL}: A Skinned Multi-Person Linear Model}.
\newblock \bibinfo{journal}{\emph{ACM Trans. Graphics (Proc. SIGGRAPH Asia)}}
  \bibinfo{volume}{34}, \bibinfo{number}{6} (\bibinfo{date}{Oct.}
  \bibinfo{year}{2015}), \bibinfo{pages}{248:1--248:16}.
\newblock


\bibitem[\protect\citeauthoryear{Lu, Mok, and Jin}{Lu et~al\mbox{.}}{2017}]%
        {lu2017new}
\bibfield{author}{\bibinfo{person}{Shufang Lu}, \bibinfo{person}{Pik~Yin Mok},
  {and} \bibinfo{person}{Xiaogang Jin}.} \bibinfo{year}{2017}\natexlab{}.
\newblock \showarticletitle{A new design concept: 3D to 2D textile pattern
  design for garments}.
\newblock \bibinfo{journal}{\emph{Computer-Aided Design}}  \bibinfo{volume}{89}
  (\bibinfo{year}{2017}), \bibinfo{pages}{35--49}.
\newblock


\bibitem[\protect\citeauthoryear{Meng, Mok, and Jin}{Meng
  et~al\mbox{.}}{2012a}]%
        {meng2012computer}
\bibfield{author}{\bibinfo{person}{Yuwei Meng}, \bibinfo{person}{Pik~Yin Mok},
  {and} \bibinfo{person}{Xiaogang Jin}.} \bibinfo{year}{2012}\natexlab{a}.
\newblock \showarticletitle{Computer aided clothing pattern design with 3D
  editing and pattern alteration}.
\newblock \bibinfo{journal}{\emph{Computer-Aided Design}} \bibinfo{volume}{44},
  \bibinfo{number}{8} (\bibinfo{year}{2012}), \bibinfo{pages}{721--734}.
\newblock


\bibitem[\protect\citeauthoryear{Meng, Wang, and Jin}{Meng
  et~al\mbox{.}}{2012b}]%
        {meng2012flexible}
\bibfield{author}{\bibinfo{person}{Yuwei Meng}, \bibinfo{person}{Charlie~CL
  Wang}, {and} \bibinfo{person}{Xiaogang Jin}.}
  \bibinfo{year}{2012}\natexlab{b}.
\newblock \showarticletitle{Flexible shape control for automatic resizing of
  apparel products}.
\newblock \bibinfo{journal}{\emph{Computer-Aided Design}} \bibinfo{volume}{44},
  \bibinfo{number}{1} (\bibinfo{year}{2012}), \bibinfo{pages}{68--76}.
\newblock


\bibitem[\protect\citeauthoryear{Meshcapade}{Meshcapade}{2020}]%
        {meshcapade}
\bibfield{author}{\bibinfo{person}{Meshcapade}.}
  \bibinfo{year}{2020}\natexlab{}.
\newblock \bibinfo{title}{meshcapade.com}.
\newblock \bibinfo{howpublished}{\url{https://meshcapade.com}}.
\newblock


\bibitem[\protect\citeauthoryear{Miguel, Bradley, Thomaszewski, Bickel,
  Matusik, Otaduy, and Marschner}{Miguel et~al\mbox{.}}{2012}]%
        {miguel2012data}
\bibfield{author}{\bibinfo{person}{Eder Miguel}, \bibinfo{person}{Derek
  Bradley}, \bibinfo{person}{Bernhard Thomaszewski}, \bibinfo{person}{Bernd
  Bickel}, \bibinfo{person}{Wojciech Matusik}, \bibinfo{person}{Miguel~A
  Otaduy}, {and} \bibinfo{person}{Steve Marschner}.}
  \bibinfo{year}{2012}\natexlab{}.
\newblock \showarticletitle{Data-driven estimation of cloth simulation models}.
  In \bibinfo{booktitle}{\emph{Computer Graphics Forum}},
  Vol.~\bibinfo{volume}{31}.
\newblock


\bibitem[\protect\citeauthoryear{Montes, Thomaszewski, Mudur, and Popa}{Montes
  et~al\mbox{.}}{2020}]%
        {montes2020computational}
\bibfield{author}{\bibinfo{person}{Juan Montes}, \bibinfo{person}{Bernhard
  Thomaszewski}, \bibinfo{person}{Sudhir Mudur}, {and} \bibinfo{person}{Tiberiu
  Popa}.} \bibinfo{year}{2020}\natexlab{}.
\newblock \showarticletitle{Computational design of skintight clothing}.
\newblock \bibinfo{journal}{\emph{ACM Trans.\ Graph.}} \bibinfo{volume}{39},
  \bibinfo{number}{4} (\bibinfo{year}{2020}), \bibinfo{pages}{105--1}.
\newblock


\bibitem[\protect\citeauthoryear{Morlet, Opsomer, Herrmann, Balmond, Gillet,
  and Fuchs}{Morlet et~al\mbox{.}}{2017}]%
        {morlet2017new}
\bibfield{author}{\bibinfo{person}{A Morlet}, \bibinfo{person}{R Opsomer},
  \bibinfo{person}{S Herrmann}, \bibinfo{person}{L Balmond}, \bibinfo{person}{C
  Gillet}, {and} \bibinfo{person}{L Fuchs}.} \bibinfo{year}{2017}\natexlab{}.
\newblock \showarticletitle{A new textiles economy: redesigning fashion’s
  future}.
\newblock \bibinfo{journal}{\emph{Ellen MacArthur Foundation}}
  (\bibinfo{year}{2017}).
\newblock


\bibitem[\protect\citeauthoryear{Narain, Samii, and O'brien}{Narain
  et~al\mbox{.}}{2012}]%
        {narain2012adaptive}
\bibfield{author}{\bibinfo{person}{Rahul Narain}, \bibinfo{person}{Armin
  Samii}, {and} \bibinfo{person}{James~F O'brien}.}
  \bibinfo{year}{2012}\natexlab{}.
\newblock \showarticletitle{Adaptive anisotropic remeshing for cloth
  simulation}.
\newblock \bibinfo{journal}{\emph{ACM Trans.\ Graph.}} \bibinfo{volume}{31},
  \bibinfo{number}{6} (\bibinfo{year}{2012}), \bibinfo{pages}{1--10}.
\newblock


\bibitem[\protect\citeauthoryear{Narayanan, Wu, Yuksel, and McCann}{Narayanan
  et~al\mbox{.}}{2019}]%
        {narayanan2019visual}
\bibfield{author}{\bibinfo{person}{Vidya Narayanan}, \bibinfo{person}{Kui Wu},
  \bibinfo{person}{Cem Yuksel}, {and} \bibinfo{person}{James McCann}.}
  \bibinfo{year}{2019}\natexlab{}.
\newblock \showarticletitle{Visual knitting machine programming}.
\newblock \bibinfo{journal}{\emph{ACM Trans.\ Graph.}} \bibinfo{volume}{38},
  \bibinfo{number}{4} (\bibinfo{year}{2019}), \bibinfo{pages}{1--13}.
\newblock


\bibitem[\protect\citeauthoryear{Nayak and Padhye}{Nayak and Padhye}{2017}]%
        {nayak:2017:automation}
\bibfield{author}{\bibinfo{person}{Rajkishore Nayak} {and}
  \bibinfo{person}{Rajiv Padhye}.} \bibinfo{year}{2017}\natexlab{}.
\newblock \bibinfo{booktitle}{\emph{Automation in Garment Manufacturing}}.
\newblock \bibinfo{publisher}{Woodhead Publishing}.
\newblock


\bibitem[\protect\citeauthoryear{Occipital}{Occipital}{2020}]%
        {occipital}
\bibfield{author}{\bibinfo{person}{Occipital}.}
  \bibinfo{year}{2020}\natexlab{}.
\newblock \bibinfo{title}{structure.io}.
\newblock \bibinfo{howpublished}{\url{https://structure.io/structure-sensor}}.
\newblock


\bibitem[\protect\citeauthoryear{Optitex}{Optitex}{2020}]%
        {optitex}
\bibfield{author}{\bibinfo{person}{Optitex}.} \bibinfo{year}{2020}\natexlab{}.
\newblock \bibinfo{title}{optitex.com}.
\newblock \bibinfo{howpublished}{\url{https://optitex.com}}.
\newblock


\bibitem[\protect\citeauthoryear{Osman, Bolkart, and Black}{Osman
  et~al\mbox{.}}{2020}]%
        {osman2020star}
\bibfield{author}{\bibinfo{person}{Ahmed~AA Osman}, \bibinfo{person}{Timo
  Bolkart}, {and} \bibinfo{person}{Michael~J Black}.}
  \bibinfo{year}{2020}\natexlab{}.
\newblock \showarticletitle{STAR: Sparse Trained Articulated Human Body
  Regressor}.
\newblock \bibinfo{journal}{\emph{arXiv preprint arXiv:2008.08535}}
  (\bibinfo{year}{2020}).
\newblock


\bibitem[\protect\citeauthoryear{Pietroni, Dumery, Falque, Liu, Vidal-Calleja,
  and Sorkine-Hornung}{Pietroni et~al\mbox{.}}{2022}]%
        {pietroni2022computational}
\bibfield{author}{\bibinfo{person}{Nico Pietroni}, \bibinfo{person}{Corentin
  Dumery}, \bibinfo{person}{Raphael Falque}, \bibinfo{person}{Mark Liu},
  \bibinfo{person}{Teresa Vidal-Calleja}, {and} \bibinfo{person}{Olga
  Sorkine-Hornung}.} \bibinfo{year}{2022}\natexlab{}.
\newblock \showarticletitle{Computational pattern making from 3D garment
  models}.
\newblock \bibinfo{journal}{\emph{ACM Transactions on Graphics (TOG)}}
  \bibinfo{volume}{41}, \bibinfo{number}{4} (\bibinfo{year}{2022}),
  \bibinfo{pages}{1--14}.
\newblock


\bibitem[\protect\citeauthoryear{Pons-Moll, Pujades, Hu, and Black}{Pons-Moll
  et~al\mbox{.}}{2017}]%
        {pons2017clothcap}
\bibfield{author}{\bibinfo{person}{Gerard Pons-Moll}, \bibinfo{person}{Sergi
  Pujades}, \bibinfo{person}{Sonny Hu}, {and} \bibinfo{person}{Michael~J
  Black}.} \bibinfo{year}{2017}\natexlab{}.
\newblock \showarticletitle{ClothCap: Seamless 4D clothing capture and
  retargeting}.
\newblock \bibinfo{journal}{\emph{ACM Transactions on Graphics (TOG)}}
  \bibinfo{volume}{36}, \bibinfo{number}{4} (\bibinfo{year}{2017}),
  \bibinfo{pages}{1--15}.
\newblock


\bibitem[\protect\citeauthoryear{Pons-Moll, Romero, Mahmood, and
  Black}{Pons-Moll et~al\mbox{.}}{2015}]%
        {Dyna:SIGGRAPH:2015}
\bibfield{author}{\bibinfo{person}{Gerard Pons-Moll}, \bibinfo{person}{Javier
  Romero}, \bibinfo{person}{Naureen Mahmood}, {and} \bibinfo{person}{Michael~J.
  Black}.} \bibinfo{year}{2015}\natexlab{}.
\newblock \showarticletitle{Dyna: A Model of Dynamic Human Shape in Motion}.
\newblock \bibinfo{journal}{\emph{ACM Trans.\ Graph.}} \bibinfo{volume}{34},
  \bibinfo{number}{4} (\bibinfo{year}{2015}).
\newblock


\bibitem[\protect\citeauthoryear{Robson, Maharik, Sheffer, and Carr}{Robson
  et~al\mbox{.}}{2011}]%
        {Robson:ContextAwareGarments:2011}
\bibfield{author}{\bibinfo{person}{C. Robson}, \bibinfo{person}{R. Maharik},
  \bibinfo{person}{A. Sheffer}, {and} \bibinfo{person}{N. Carr}.}
  \bibinfo{year}{2011}\natexlab{}.
\newblock \showarticletitle{Context-Aware Garment Modeling from Sketches}.
\newblock \bibinfo{journal}{\emph{Computers \& Graphics (Proc. SMI 2011)}}
  \bibinfo{volume}{35}, \bibinfo{number}{3} (\bibinfo{year}{2011}),
  \bibinfo{pages}{604--613}.
\newblock


\bibitem[\protect\citeauthoryear{Rose, Sheffer, Wither, Cani, and Thibert}{Rose
  et~al\mbox{.}}{2007}]%
        {Rose::DevelopableSurfaces:2007}
\bibfield{author}{\bibinfo{person}{Kenneth Rose}, \bibinfo{person}{Alla
  Sheffer}, \bibinfo{person}{Jamie Wither}, \bibinfo{person}{Marie-Paule Cani},
  {and} \bibinfo{person}{Boris Thibert}.} \bibinfo{year}{2007}\natexlab{}.
\newblock \showarticletitle{Developable surfaces from arbitrary sketched
  boundaries}. In \bibinfo{booktitle}{\emph{Proc.\ Symposium on Geometry
  Processing}}. Eurographics Association, \bibinfo{pages}{163--172}.
\newblock


\bibitem[\protect\citeauthoryear{Sharp and Crane}{Sharp and Crane}{2018}]%
        {sharp2018variational}
\bibfield{author}{\bibinfo{person}{Nicholas Sharp} {and}
  \bibinfo{person}{Keenan Crane}.} \bibinfo{year}{2018}\natexlab{}.
\newblock \showarticletitle{Variational surface cutting}.
\newblock \bibinfo{journal}{\emph{ACM Trans.\ Graph.}} \bibinfo{volume}{37},
  \bibinfo{number}{4} (\bibinfo{year}{2018}), \bibinfo{pages}{1--13}.
\newblock


\bibitem[\protect\citeauthoryear{Sieger and Botsch}{Sieger and Botsch}{2020}]%
        {pmp-library}
\bibfield{author}{\bibinfo{person}{Daniel Sieger} {and} \bibinfo{person}{Mario
  Botsch}.} \bibinfo{year}{2020}\natexlab{}.
\newblock \bibinfo{title}{The Polygon Mesh Processing Library}.
\newblock
\newblock
\newblock
\shownote{http://www.pmp-library.org.}


\bibitem[\protect\citeauthoryear{SizeGermany}{SizeGermany}{2020}]%
        {SizeGermany}
\bibfield{author}{\bibinfo{person}{SizeGermany}.}
  \bibinfo{year}{2020}\natexlab{}.
\newblock \bibinfo{title}{SizeGermany}.
\newblock \bibinfo{howpublished}{\url{https://portal.sizegermany.de}}.
\newblock


\bibitem[\protect\citeauthoryear{Sorkine and Alexa}{Sorkine and Alexa}{2007}]%
        {ARAP_modeling:2007}
\bibfield{author}{\bibinfo{person}{Olga Sorkine} {and} \bibinfo{person}{Marc
  Alexa}.} \bibinfo{year}{2007}\natexlab{}.
\newblock \showarticletitle{As-Rigid-As-Possible Surface Modeling}. In
  \bibinfo{booktitle}{\emph{Proc.\ Symposium on Geometry Processing}}.
  \bibinfo{pages}{109--116}.
\newblock


\bibitem[\protect\citeauthoryear{Sumner and Popovi{\'c}}{Sumner and
  Popovi{\'c}}{2004}]%
        {sumner2004deformation}
\bibfield{author}{\bibinfo{person}{Robert~W Sumner} {and}
  \bibinfo{person}{Jovan Popovi{\'c}}.} \bibinfo{year}{2004}\natexlab{}.
\newblock \showarticletitle{Deformation transfer for triangle meshes}.
\newblock \bibinfo{journal}{\emph{ACM Trans.\ Graph.}} \bibinfo{volume}{23},
  \bibinfo{number}{3} (\bibinfo{year}{2004}), \bibinfo{pages}{399--405}.
\newblock


\bibitem[\protect\citeauthoryear{Tamstorf and Grinspun}{Tamstorf and
  Grinspun}{2013}]%
        {tamstorf2013discrete}
\bibfield{author}{\bibinfo{person}{Rasmus Tamstorf} {and}
  \bibinfo{person}{Eitan Grinspun}.} \bibinfo{year}{2013}\natexlab{}.
\newblock \showarticletitle{Discrete bending forces and their Jacobians}.
\newblock \bibinfo{journal}{\emph{Graphical models}} \bibinfo{volume}{75},
  \bibinfo{number}{6} (\bibinfo{year}{2013}), \bibinfo{pages}{362--370}.
\newblock


\bibitem[\protect\citeauthoryear{Tang, Wang, Liu, Tong, and Manocha}{Tang
  et~al\mbox{.}}{2018}]%
        {tang2018cloth}
\bibfield{author}{\bibinfo{person}{Min Tang}, \bibinfo{person}{Tongtong Wang},
  \bibinfo{person}{Zhongyuan Liu}, \bibinfo{person}{Ruofeng Tong}, {and}
  \bibinfo{person}{Dinesh Manocha}.} \bibinfo{year}{2018}\natexlab{}.
\newblock \showarticletitle{I-cloth: incremental collision handling for
  GPU-based interactive cloth simulation}.
\newblock \bibinfo{journal}{\emph{ACM Trans.\ Graph.}} \bibinfo{volume}{37},
  \bibinfo{number}{6} (\bibinfo{year}{2018}), \bibinfo{pages}{1--10}.
\newblock


\bibitem[\protect\citeauthoryear{Turquin, Wither, Boissieux, Cani, and
  Hughes}{Turquin et~al\mbox{.}}{2007}]%
        {Turquin:SketchInterface:2007}
\bibfield{author}{\bibinfo{person}{Emmanuel Turquin}, \bibinfo{person}{Jamie
  Wither}, \bibinfo{person}{Laurence Boissieux}, \bibinfo{person}{Marie-Paule
  Cani}, {and} \bibinfo{person}{John~F Hughes}.}
  \bibinfo{year}{2007}\natexlab{}.
\newblock \showarticletitle{A sketch-based interface for clothing virtual
  characters}.
\newblock \bibinfo{journal}{\emph{IEEE Computer Graphics and Applications}}
  \bibinfo{volume}{27}, \bibinfo{number}{1} (\bibinfo{year}{2007}).
\newblock


\bibitem[\protect\citeauthoryear{Umetani, Kaufman, Igarashi, and
  Grinspun}{Umetani et~al\mbox{.}}{2011}]%
        {Umetani:SensitiveCouture:2011}
\bibfield{author}{\bibinfo{person}{Nobuyuki Umetani}, \bibinfo{person}{Danny~M
  Kaufman}, \bibinfo{person}{Takeo Igarashi}, {and} \bibinfo{person}{Eitan
  Grinspun}.} \bibinfo{year}{2011}\natexlab{}.
\newblock \showarticletitle{Sensitive couture for interactive garment modeling
  and editing.}
\newblock \bibinfo{journal}{\emph{ACM Trans. Graph.}} \bibinfo{volume}{30},
  \bibinfo{number}{4} (\bibinfo{year}{2011}), \bibinfo{pages}{90--1}.
\newblock


\bibitem[\protect\citeauthoryear{Volino, Cordier, and Magnenat-Thalmann}{Volino
  et~al\mbox{.}}{2005}]%
        {Volino:EarlyGarmentDesign:2005}
\bibfield{author}{\bibinfo{person}{Pascal Volino}, \bibinfo{person}{Frederic
  Cordier}, {and} \bibinfo{person}{Nadia Magnenat-Thalmann}.}
  \bibinfo{year}{2005}\natexlab{}.
\newblock \showarticletitle{From early virtual garment simulation to
  interactive fashion design}.
\newblock \bibinfo{journal}{\emph{Computer Aided Design}} \bibinfo{volume}{37},
  \bibinfo{number}{6} (\bibinfo{year}{2005}), \bibinfo{pages}{593--608}.
\newblock


\bibitem[\protect\citeauthoryear{Wang, Wang, and Yuen}{Wang
  et~al\mbox{.}}{2005}]%
        {wang2005design}
\bibfield{author}{\bibinfo{person}{Charlie~CL Wang}, \bibinfo{person}{Yu Wang},
  {and} \bibinfo{person}{Matthew~MF Yuen}.} \bibinfo{year}{2005}\natexlab{}.
\newblock \showarticletitle{Design automation for customized apparel products}.
\newblock \bibinfo{journal}{\emph{Computer-aided design}} \bibinfo{volume}{37},
  \bibinfo{number}{7} (\bibinfo{year}{2005}), \bibinfo{pages}{675--691}.
\newblock


\bibitem[\protect\citeauthoryear{Wang}{Wang}{2018}]%
        {wang2018rule}
\bibfield{author}{\bibinfo{person}{Huamin Wang}.}
  \bibinfo{year}{2018}\natexlab{}.
\newblock \showarticletitle{Rule-free sewing pattern adjustment with precision
  and efficiency}.
\newblock \bibinfo{journal}{\emph{ACM Trans.\ Graph.}} \bibinfo{volume}{37},
  \bibinfo{number}{4} (\bibinfo{year}{2018}), \bibinfo{pages}{1--13}.
\newblock


\bibitem[\protect\citeauthoryear{Wang, O'Brien, and Ramamoorthi}{Wang
  et~al\mbox{.}}{2011}]%
        {wang2011data}
\bibfield{author}{\bibinfo{person}{Huamin Wang}, \bibinfo{person}{James~F
  O'Brien}, {and} \bibinfo{person}{Ravi Ramamoorthi}.}
  \bibinfo{year}{2011}\natexlab{}.
\newblock \showarticletitle{Data-driven elastic models for cloth: modeling and
  measurement}.
\newblock \bibinfo{journal}{\emph{ACM Trans.\ Graph.}} \bibinfo{volume}{30},
  \bibinfo{number}{4} (\bibinfo{year}{2011}), \bibinfo{pages}{1--12}.
\newblock


\bibitem[\protect\citeauthoryear{Wang, Ceylan, Popovic, and Mitra}{Wang
  et~al\mbox{.}}{2018}]%
        {Wang:GarmentShapeSpace:2018}
\bibfield{author}{\bibinfo{person}{Tuanfeng~Y. Wang}, \bibinfo{person}{Duygu
  Ceylan}, \bibinfo{person}{Jovan Popovic}, {and} \bibinfo{person}{Niloy~J.
  Mitra}.} \bibinfo{year}{2018}\natexlab{}.
\newblock \showarticletitle{Learning a Shared Shape Space for Multimodal
  Garment Design}.
\newblock \bibinfo{journal}{\emph{ACM Trans.\ Graph.}} \bibinfo{volume}{37},
  \bibinfo{number}{6} (\bibinfo{year}{2018}), \bibinfo{pages}{1:1--1:14}.
\newblock
\urldef\tempurl%
\url{https://doi.org/10.1145/3272127.3275074}
\showDOI{\tempurl}


\bibitem[\protect\citeauthoryear{Wibowo, Sakamoto, Mitani, and Igarashi}{Wibowo
  et~al\mbox{.}}{2012}]%
        {Wibowo:Dressup:2012}
\bibfield{author}{\bibinfo{person}{Amy Wibowo}, \bibinfo{person}{Daisuke
  Sakamoto}, \bibinfo{person}{Jun Mitani}, {and} \bibinfo{person}{Takeo
  Igarashi}.} \bibinfo{year}{2012}\natexlab{}.
\newblock \showarticletitle{DressUp: a 3D interface for clothing design with a
  physical mannequin}. In \bibinfo{booktitle}{\emph{Proc.\ International
  Conference on Tangible, Embedded and Embodied Interaction}}. ACM,
  \bibinfo{pages}{99--102}.
\newblock


\bibitem[\protect\citeauthoryear{Wolff, Herholz, and Sorkine-Hornung}{Wolff
  et~al\mbox{.}}{2019}]%
        {wolff2019reflection}
\bibfield{author}{\bibinfo{person}{Katja Wolff}, \bibinfo{person}{Philipp
  Herholz}, {and} \bibinfo{person}{Olga Sorkine-Hornung}.}
  \bibinfo{year}{2019}\natexlab{}.
\newblock \showarticletitle{Reflection Symmetry in Textured Sewing Patterns}.
\newblock  (\bibinfo{year}{2019}).
\newblock


\bibitem[\protect\citeauthoryear{Wolff and Sorkine-Hornung}{Wolff and
  Sorkine-Hornung}{2019}]%
        {wolff2019wallpaper}
\bibfield{author}{\bibinfo{person}{Katja Wolff} {and} \bibinfo{person}{Olga
  Sorkine-Hornung}.} \bibinfo{year}{2019}\natexlab{}.
\newblock \showarticletitle{Wallpaper pattern alignment along garment seams}.
\newblock \bibinfo{journal}{\emph{ACM Transactions on Graphics (TOG)}}
  \bibinfo{volume}{38}, \bibinfo{number}{4} (\bibinfo{year}{2019}),
  \bibinfo{pages}{1--12}.
\newblock


\bibitem[\protect\citeauthoryear{Wu, Swan, and Yuksel}{Wu
  et~al\mbox{.}}{2019}]%
        {wu2019knittable}
\bibfield{author}{\bibinfo{person}{Kui Wu}, \bibinfo{person}{Hannah Swan},
  {and} \bibinfo{person}{Cem Yuksel}.} \bibinfo{year}{2019}\natexlab{}.
\newblock \showarticletitle{Knittable stitch meshes}.
\newblock \bibinfo{journal}{\emph{ACM Trans.\ Graph.}} \bibinfo{volume}{38},
  \bibinfo{number}{1} (\bibinfo{year}{2019}), \bibinfo{pages}{1--13}.
\newblock


\bibitem[\protect\citeauthoryear{Xu, Zhang, Wang, and Bao}{Xu
  et~al\mbox{.}}{2006}]%
        {xu2006poisson}
\bibfield{author}{\bibinfo{person}{Dong Xu}, \bibinfo{person}{Hongxin Zhang},
  \bibinfo{person}{Qing Wang}, {and} \bibinfo{person}{Hujun Bao}.}
  \bibinfo{year}{2006}\natexlab{}.
\newblock \showarticletitle{Poisson shape interpolation}.
\newblock \bibinfo{journal}{\emph{Graphical models}} \bibinfo{volume}{68},
  \bibinfo{number}{3} (\bibinfo{year}{2006}), \bibinfo{pages}{268--281}.
\newblock


\end{thebibliography}

\appendix

\section{Stretch computation} \label{app:stretch}
To measure stretch we consider the singular value decomposition of the deformation gradient $\mathbf F \in \mathds R^{2\times 2}$. The SVD is given by
\begin{align}
    \mathbf F = \mathbf U  \begin{pmatrix} \sigma_1 & 0 \\ 0 & \sigma_2 \end{pmatrix}  \mathbf V^\top
\end{align}
with $\mathbf U, \mathbf V \in \mathds R^{2\times 2}$. The singular values $\sigma_1$ and $\sigma_2$ are called \emph{principal stretches}. If they both have the value 1, the deformation is merely a rotation and no stretch is induced. A value below 1 indicates compression, which we always allow, since we do not want to shrink the garment. A sensible measure for stretch is for example $(\sigma_1 - 1)^2 + (\sigma_2 - 1)^2$. We want to find the matrix $\bar{\mathbf F}$ that limits principal stretch to the interval $\left[0, 1 + \delta \right]$ and is at the same time as close as possible to $\mathbf F$. 

We can find it by clipping the singular values at the desired stretch interval:
\begin{align}
    \bar{\sigma}_i = \begin{cases} 1 + \delta & \sigma_i > 1 + \delta\\  \sigma_i & \text{otherwise.} \end{cases}
\end{align}
The modified deformation gradient  $\bar{ \mathbf F }$ is obtained by substituting the new singular values
\begin{align}
   \bar{ \mathbf F } = \mathbf U  \begin{pmatrix} \bar{\sigma}_1 & 0 \\ 0 & \bar{\sigma}_2 \end{pmatrix}  \mathbf V^\top.
\end{align}
To obtain the modified restshape triangle we multiply the current simulation triangle with the inverse deformation gradient
\begin{align}
   \bar{ \mathbf F }^{-1} = \mathbf U  \begin{pmatrix} 1 /\bar{\sigma}_1 & 0 \\ 0 & 1 / \bar{\sigma}_2 \end{pmatrix}  \mathbf V^\top.
\end{align}
For non-degenrate triangles, this inverse always exists. If one of the singular values is zero, we resort to the pseudo inverse.

\section{As-rigid-as-possible mesh stitching} \label{app:arap}
The rest shape triangles are modified individually and can generally not be combined to form a consistent mesh. One reason for this problem is that edges that are shared between two neighbouring triangles do not necessarily have the same length. We want to find a connected restshape mesh that maintains the shape of the triangles as much as possible. In other words, the transformation from the modified 2D triangle $\bar{P}_{\hat{\mathbf x}_{\mathbf t}}$ to the new 3D rest shape triangle $\hat{\mathbf x}_{\mathbf t}$ should be a rigid transformation. This motivates the use of as-rigid-as-possible shape deformation \cite{ARAP_modeling:2007}. We refer the reader to the paper for further details.

\end{document}